\documentclass[a4paper,12pt]{spieman}  
\usepackage{amsmath,amsfonts,amssymb}
\usepackage{graphicx}
\usepackage{multirow}
\usepackage{setspace}
\usepackage{tocloft}
\usepackage{multicol}
\def\bm#1{\mbox{\boldmath $#1$}}
\usepackage[left]{lineno} 

\title{In-Orbit Performance and Calibration of the Hard X-ray Imager (HXI) onboard Hitomi (ASTRO-H)}

\author[a,*]{Kouichi~Hagino} 
\author[b,c]{Kazuhiro~Nakazawa} 
\author[d]{Goro~Sato} 
\author[d]{Motohide~Kokubun} 
\author[e,f]{Teruaki~Enoto} 
\author[g]{Yasushi~Fukazawa} 
\author[d,h]{Katsuhiro~Hayashi} 
\author[i]{Jun~Kataoka} 
\author[g]{Junichiro~Katsuta} 
\author[j]{Shogo~B.~Kobayashi} 
\author[k,l]{Philippe~Laurent} 
\author[k]{Francois~Lebrun} 
\author[l]{Olivier~Limousin} 
\author[l]{Daniel~Maier} 
\author[m]{Kazuo~Makishima} 
\author[i]{Taketo~Mimura} 
\author[b]{Katsuma~Miyake} 
\author[g,n]{Tsunefumi~Mizuno} 
\author[d]{Kunishiro~Mori} 
\author[b]{Hiroaki~Murakami} 
\author[o]{Takeshi~Nakamori} 
\author[p]{Toshio~Nakano} 
\author[q,r]{Hirofumi~Noda} 
\author[s]{Hirokazu~Odaka} 
\author[f]{Masanori~Ohno} 
\author[d]{Masayuki~Ohta} 
\author[t]{Shinya~Saito} 
\author[d]{Rie~Sato} 
\author[u]{Hiroyasu~Tajima} 
\author[g]{Hiromitsu~Takahashi} 
\author[d]{Tadayuki~Takahashi} 
\author[v]{Shin'ichiro~Takeda} 
\author[j]{Takaaki~Tanaka} 
\author[w]{Yukikatsu~Terada} 
\author[x]{Hideki~Uchiyama} 
\author[t]{Yasunobu~Uchiyama} 
\author[d]{Shin~Watanabe} 
\author[h,u]{Kazutaka~Yamaoka} 
\author[y]{Yoichi~Yatsu} 
\author[m]{Takayuki~Yuasa} 
\author[ ]{and~the~HXI~team}

\affil[a]{Department of Physics, Tokyo University of Science, 2641 Yamazaki, Noda, Chiba, 278-8510, Japan}
\affil[b]{Department of Physics, The University of Tokyo, 7-3-1 Hongo, Bunkyo-ku, Tokyo 113-0033, Japan}
\affil[c]{Research Center for the Early Universe, School of Science, The University of Tokyo, 7-3-1 Hongo, Bunkyo-ku, Tokyo 113-0033, Japan}
\affil[d]{Japan Aerospace Exploration Agency, Institute of Space and Astronautical Science, 3-1-1 Yoshino-dai, Chuo-ku, Sagamihara, Kanagawa 252-5210, Japan}
\affil[e]{Department of Astronomy, Kyoto University, Kitashirakawa-Oiwake-cho, Sakyo-ku, Kyoto 606-8502, Japan}
\affil[f]{The Hakubi Center for Advanced Research, Kyoto University, Kyoto 606-8302, Japan}
\affil[g]{School of Science, Hiroshima University, 1-3-1 Kagamiyama, Higashi-Hiroshima 739-8526, Japan}
\affil[h]{Department of Physics, Nagoya University, Furo-cho, Chikusa-ku, Nagoya, Aichi 464-8602, Japan}
\affil[i]{Research Institute for Science and Engineering, Waseda University, 3-4-1 Ohkubo, Shinjuku, Tokyo, 169-8555, Japan}
\affil[j]{Department of Physics, Kyoto University, Kitashirakawa-Oiwake-Cho, Sakyo, Kyoto 606-8502, Japan}
\affil[k]{Laboratoire APC, 10 rue Alice Domon et L\'eonie Duquet, 75013 Paris, France}
\affil[l]{CEA Saclay, 91191 Gif sur Yvette, France}
\affil[m]{Institute of Physical and Chemical Research, 2-1 Hirosawa, Wako, Saitama 351-0198}
\affil[n]{Hiroshima Astrophysical Science Center, Hiroshima University, Higashi-Hiroshima, Hiroshima 739-8526, Japan}
\affil[o]{Faculty of Science, Yamagata University, 1-4-12 Kojirakawa-machi, Yamagata, Yamagata 990-8560, Japan}
\affil[p]{RIKEN Nishina Center, 2-1 Hirosawa, Wako, Saitama 351-0198, Japan}
\affil[q]{Frontier Research Institute for Interdisciplinary Sciences, Tohoku University,  6-3 Aramakiazaaoba, Aoba-ku, Sendai, Miyagi 980-8578, Japan}
\affil[r]{Astronomical Institute, Tohoku University, 6-3 Aramakiazaaoba, Aoba-ku, Sendai, Miyagi 980-8578, Japan}
\affil[s]{Kavli Institute for Particle Astrophysics and Cosmology, Stanford University, 452 Lomita Mall, Stanford, CA 94305, USA}
\affil[t]{Department of Physics, Rikkyo University, 3-34-1 Nishi-Ikebukuro, Toshima-ku, Tokyo 171-8501, Japan}
\affil[u]{Institute for Space-Earth Environmental Research, Nagoya University, Furo-cho, Chikusa-ku, Nagoya, Aichi 464-8601, Japan}
\affil[v]{Okinawa Institute of Science and Technology Graduate University, 1919-1 Tancha, Onna-son Okinawa, 904-0495, Japan}
\affil[w]{Department of Physics, Saitama University, 255 Shimo-Okubo, Sakura-ku, Saitama, 338-8570, Japan}
\affil[x]{Faculty of Education, Shizuoka University, 836 Ohya, Suruga-ku, Shizuoka 422-8529, Japan}
\affil[y]{Department of Physics, Tokyo Institute of Technology, 2-12-1 Ookayama, Meguro-ku, Tokyo 152-8550, Japan}

\cftpagenumbersoff{figure}
\cftpagenumbersoff{table} 
\begin{document} 
\maketitle


\begin{abstract}
The Hard X-ray Imager (HXI) onboard {\it Hitomi} (ASTRO-H) is an imaging spectrometer covering hard X-ray energies of 5--80~keV. Combined with the hard X-ray telescope, it enables imaging spectroscopy with an angular resolution of $1^\prime.7$ half-power diameter, in a field of view of $9^\prime\times9^\prime$. The main imager is composed of 4 layers of Si detectors and 1 layer of CdTe detector, stacked to cover wide energy band up to 80~keV, surrounded by an active shield made of BGO scintillator to reduce the background. The HXI started observations 12 days before the {\it Hitomi} loss, and successfully obtained data from G21.5$-$0.9, Crab and blank sky. Utilizing these data, we calibrate the detector response and study properties of in-orbit background. The observed Crab spectra agree well with a powerlaw model convolved with the detector response, within 5\% accuracy. We find that albedo electrons in specified orbit strongly affect the background of Si top layer, and establish a screening method to reduce it. The background level over the full field of view after all the processing and screening is as low as the pre-flight requirement of $1\textrm{--}3\times10^{-4}$~counts~s$^{-1}$~cm$^{-2}$~keV$^{-1}$.
\end{abstract}


{\noindent \footnotesize\textbf{*}Kouichi Hagino,  \linkable{hagino@rs.tus.ac.jp} }

\begin{spacing}{2}   

\section{Introduction} \label{sect:intro}
An international X-ray satellite, {\it Hitomi}, led by Japan was launched on 2016 February 17 by an H-IIA rocket at the Tanegashima Space Center in Japan, and placed in a low-Earth orbit with an altitude of 575~km and an inclination angle of $31^\circ$\cite{Takahashi2017}. {\it Hitomi} carries four types of instruments covering a wide energy range from soft X-ray to soft Gamma-ray. The hard X-ray imaging system composed of two sets of the Hard X-ray Imagers (HXI)\cite{Nakazawa2017} and two sets of the Hard X-ray Telescopes (HXT)\cite{Matsumoto2017} is capable of imaging spectroscopy in the hard X-ray band ranging from 5~keV to 80~keV\cite{Nakazawa2017}. The two HXI systems are referred to as HXI1 and HXI2, individually paired with HXT1 and HXT2, respectively. Thanks to the focusing optics, the sensitivity of the hard X-ray imaging system for the point source is 100 times better than those of non-focusing instruments in the hard X-ray bands, such as {\it Suzaku}/HXD\cite{Takahashi2006}. 

The HXI is composed of a stacked semiconductor imager\cite{Watanabe2007a,Kokubun2010,Odaka2012,Sato2016} and active shields surrounding the imager. The imager consists of 5 layers of double-sided strip detectors with a strip pitch of 250~$\mu$m and detector area of $32\times32$~mm$^2$. Upper 4 layers are the double-sided Si strips detectors (DSSDs) with a thickness of 500~$\mu$m\cite{Takeda2007,Nakazawa2007,Hayashi2013}, and the bottom layer is the CdTe double-sided strip detector (CdTe-DSD) with a thickness of 750~$\mu$m\cite{Takahashi2001,Ishikawa2008,Watanabe2009,Ishikawa2010,Hagino2012}. The DSSDs have p- and n-type strips on the surface of the top and bottom sides of an n-type Si wafer, while the CdTe-DSDs have Pt- and Al-strips on those of a p-type CdTe wafer. By applying positive bias voltages to the n-side of the DSSDs and the Al-side of the CdTe-DSDs, holes and electrons generated by the incident X-ray photon are collected by the p-side/Pt-side strips and the n-side/Al-side strips, respectively. The X-ray induced charge on the strip electrodes are read out utilizing dedicated low-noise front-end ASICs (application specific integrated circuit)\cite{Tajima2004}, which are connected to the individual strips. The active shields consist of nine BGO (Bi$_4$Ge$_3$O$_{12}$) scintillators, arranged as a well-type structure. Their thicknesses are typically $\simeq3$~cm in order to stop protons with energy $\lesssim100$~MeV, trapped at the South Atlantic Anomaly (SAA). The scintillation light of each BGO is read-out by an avalanche photo-diode (APD)\cite{Saito2013}. By processing the read-out signals from the APDs in digital filters, veto signals are generated and used for reducing the detector background\cite{Ohno2016}.

Although the HXI was lost\cite{Nakazawa2017}, thorough investigations and evaluations of its in-orbit performance is of great importance for planning and designing future hard X-ray missions. In this paper, we describe the in-orbit performance and calibration results of the HXI. In section~\ref{sect:ope}, in-orbit operations and functionalities are summarized. Standard analysis method of the HXI are described in section~\ref{sect:ana}. Detailed performances on the non-X-ray background and energy response are presented in sections~\ref{sect:resp} and \ref{sect:nxb}.

\section{In-flight Operations}  \label{sect:ope}
\subsection{Initial Operations and Observations}
After the deployment of the extensible optical bench on 2016 February 28, the temperature of the HXI was gradually cooled to the operation temperature of $-25^\circ$C. On 2016 March 8, a start-up operation of the HXI started. High voltages of the APDs and the DSSD/CdTe-DSD were applied one by one, and reached the optimum values on March 12 for HXI1 and March 14 for HXI2.

After turned on, the HXI performed several observations as listed in Tab.~\ref{tab:obs}. In spite of the short lifetime of the HXI, X-ray photons from 3 astronomical objects (IGR J16318$-$4848, G21.5$-$0.9 and Crab nebula) were successfully detected. As well as these data, the HXI observed 164.3~ks (HXI1) and 163.8~ks (HXI2) of blank sky data and 158.7~ks (HXI1) and 160.5~ks (HXI2) of  Earth occultation data including both bright and night Earth. According to the hard X-ray observations by {\it Swift}\cite{Ajello2008}, count rate on the HXI due to the albedo X-ray/gamma-rays is estimated to be less than $\sim10^{-5}$~counts~s$^{-1}$~keV$^{-1}$~cm$^{-2}$. Since this count rate is negligible compared with the background rate as shown in section~\ref{sect:nxb}, the Earth occultation data is referred to as the non-X-ray background (NXB) in this paper. On the other hand, the summed data of observation sequences named ``None2'', ``IRU check out'' and ``RXJ1856.5$-$3754'', which includes the cosmic X-ray background, is referred to as blank sky. These background data provide fruitful information on the in-orbit background properties in the hard X-ray energies as described in section~\ref{sect:nxb}.

\begin{table*}[tbp]
\caption{Observation log of the HXI}
\begin{center}
\begin{tabular}{lllll}
\hline
Start time & Stop time & OBSID & Target name & Notes \\
\hline
03-11 21:24 & 03-13 17:56 & 10042020--10042030 & IGR J16318$-$4848 & HXI1 on\\
03-13 17:56 & 03-14 16:20 & 100042040 & IGR J16318$-$4848 & Stray light\\
03-14 16:20 & 03-14 18:00 & 000007010 & None2 &\\
03-14 18:00 & 03-15 17:56 & 000007020 & None2 & HXI2 on\\
03-15 17:56 & 03-16 19:40 & 000008010--000008060 & IRU Check out & \\
03-16 19:40 & 03-19 17:00 & 100043010--100043040 & RXJ1856.5$-$3754 & \\
03-19 17:00 & 03-23 13:30 & 100050010--100050050 & G21.5$-$0.9 & \\
03-23 13:30 & 03-24 11:22 & 100043050 & RXJ1856.5$-$3754 & \\
03-24 11:22 & 03-25 11:28 & 100043060 & RXJ1856.5$-$3754 & DTHR* changed\\
03-25 11:28 & 03-25 18:01 & 000007010--000007020 & Crab & \\
\hline
\end{tabular}
\end{center}
\begin{flushleft}
\footnotesize
* DTHR is the ASIC ADC digital thresholds for reducing the data size.
\end{flushleft}
\label{tab:obs}
\end{table*}%

\subsection{Basic Characteristics}
In orbit, all the basic functions of the HXI worked properly. Here, functionalities of the HXI in flight are briefly summarized. For more details, please refer to Nakazawa et al. (2018)\cite{Nakazawa2017}.
In the imagers, there was no damage or degradation due to the launch. All read-out channels of all the ASICs worked properly, noise levels were consistent with the ground calibration. The energy resolution was evaluated by fitting the on-board calibration source spectra, which was mounted just above the top layer of the DSSD. From these data, good energy resolutions of 1.0~keV at 13.9~keV and 2.0~keV at 59.5~keV in full-width at half-maximum (FWHM) were obtained. Also, the energy gain was very stable within an uncertainty of less than 1~bin of the pulse-height invariant (PI) at 59.5 keV, corresponding to 0.1~keV or $\sim0.2$\%.

The active shields also showed good performances. Low energy threshold of each BGO scintillator was the same as the ground calibration results, and anti-coincidence rate was consistent with the pre-launch estimation. Light curves of the veto signals from the active shields clearly showed variability corresponding to the geomagnetic cut-off rigidity and decay of the activation component after passages of the SAA. It indicated that the active shields properly monitored the variability of cosmic-ray in the geomagnetic cut-off rigidity and SAA.

\section{Analysis Method} \label{sect:ana}
In this section, the standard analysis method for the HXI data is summarized. It is composed of four steps: gain correction, event reconstruction, screening and dead-time correction.

\subsection{Gain Correction}
In the gain correction process, gain-corrected EPI (pulse height invariant in units of keV, in real number) is calculated. The EPI is generated from the raw ADC value in the gain correction process, and used only in the event reconstruction process. After the event reconstruction, the EPI is converted into PI (pulse height invariant in integer number), which is used in further screening and scientific analysis. The relation between PI and EPI is expressed as ${\rm PI}/10 < {\rm EPI} < ({\rm PI}+1)/10$.
For example, ${\rm EPI}=23.17$~keV (Cd K$\alpha_1$ line) is converted to ${\rm PI}=231$.

At the first step of the gain correction, all signals from bad channels are excluded from the following processes. In the flight models, only strips located at the edge of the detector, where the leakage current is higher than the other strips, are defined as bad channels. Thus, the detector area within $31.5\times 31.5$~mm$^2$ (126~strips$\times$126~strips) is available for imaging spectroscopy which corresponds to $9^\prime.03\times9^\prime.03$. Then, the common mode noise is subtracted from ADC values, before correcting the gain. The common mode noise is a noise where all channels in one readout ASIC coherently fluctuate. It is estimated in the ASICs by recording the median ADC value, which is 16th smallest in all channels in one ASIC. By the common mode subtraction, the pedestal level of each channel is corrected to zero. Finally, the ADC values of good strips are converted to EPI with third order polynomial functions.

The gain-correction functions are determined based on the ground calibrations conducted in December and October 2014 for HXI1 and HXI2, respectively. The two HXIs were operated in a low-temperature chamber at ISAS, where X-ray/gamma-ray photons from radio-active isotopes  $^{241}$Am, $^{133}$Ba, $^{57}$Co and $^{55}$Fe irradiated the instrument. From these data, the correspondence between ADC values and photon energies for X-ray and gamma-ray lines are obtained and listed in Tab.~\ref{tab:lines}. Between these lines, the ADC-energy correspondance are interpolated with third order spline functions, which are used as the gain-correction function.

\begin{table}[tbp]
\caption{X-ray and gamma-ray lines used for gain calibration}
\begin{center}
\begin{tabular}{lll}
\hline
Layer & Side** & Lines (keV)\\
\hline
0 & Top & 5.9, 13.9, 17.8, 20.8, 26.3, 30.8, 35.0, 59.5, 81.0, 122*\\
 & Bottom & 5.9, 30.8, 59.5, 81.0*, 122*\\
\hline
1 & Top &13.9, 17.8, 20.8, 26.3, 30.8, 35.0, 59.5, 81.0, 122*\\
 & Bottom & 30.8, 59.5, 81.0*, 122*\\
\hline
2 & Top & 13.9, 17.8, 20.8, 26.3, 30.8, 35.0, 59.5, 81.0, 122*\\
 & Bottom & 30.8, 59.5, 81.0*, 122*\\
\hline
3 & Top & 13.9*, 17.8, 20.8, 30.8, 35.0, 59.5, 81.0, 122*\\
 & Bottom & 30.8, 59.5, 81.0*, 122*\\
\hline
4 & Top & 17.8*, 30.8, 35.0, 59.5, 81.0, 122\\
 & Bottom & 17.8*, 30.8, 35.0, 59.5, 81.0, 122, 136*\\
\hline
\end{tabular}
\end{center}
\begin{flushleft}
\footnotesize
* Spectra from sum of all the channels in one ASIC are used for gain calibration, because of lack of the number of photons.\\
** Top and bottom sides of the DSSD are the p- and n-side, and those of the CdTe-DSD are Pt- and Al-side, respectively.
\end{flushleft}
\label{tab:lines}
\end{table}%

\subsection{Event Reconstruction}
\begin{figure}[tbp]
\begin{center}
\includegraphics[width=0.7\hsize]{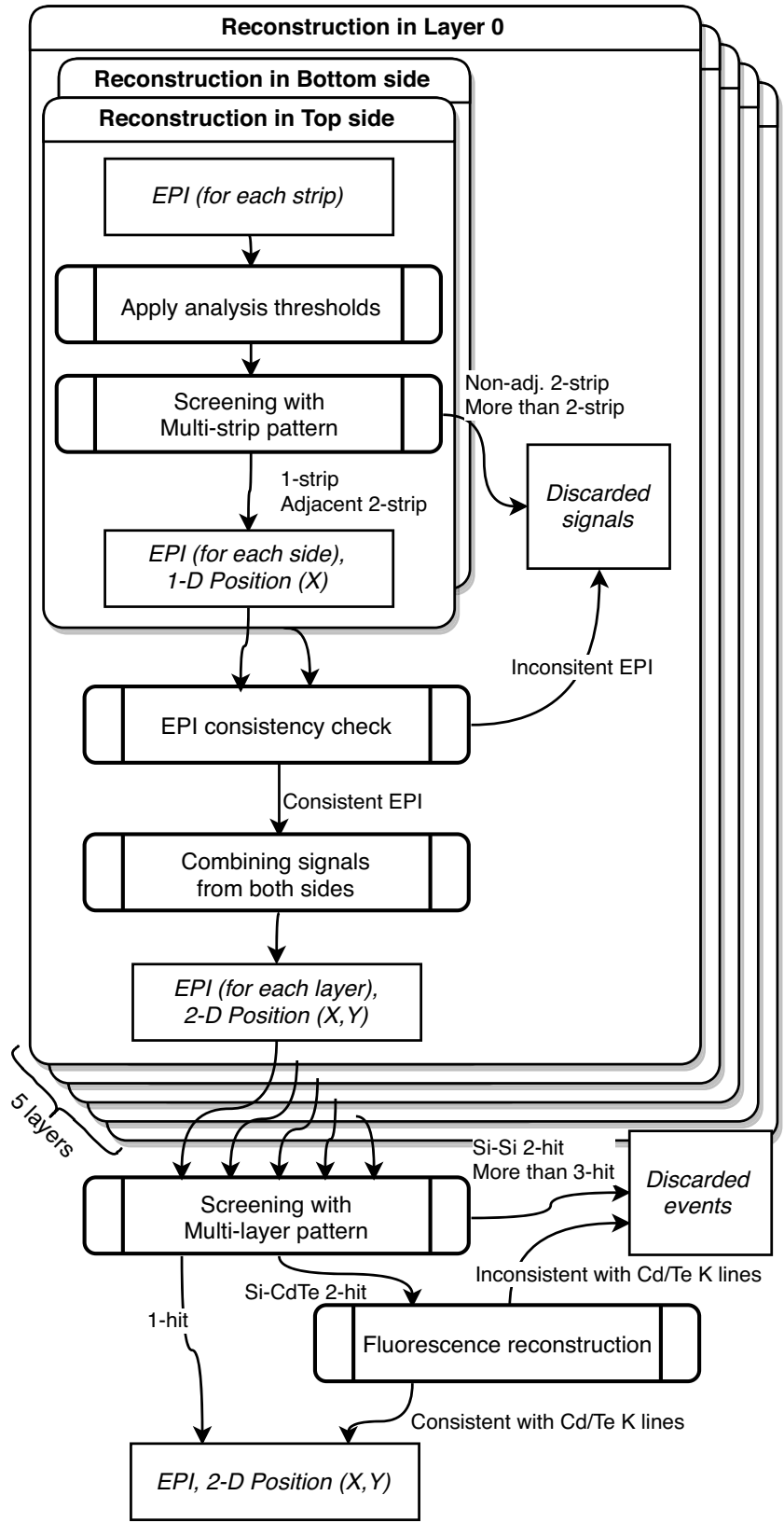}
\caption{Overview of the event reconstruction of the HXI.}
\label{fig:reconstdia}
\end{center}
\end{figure}
In order to obtain photon information from the gain-corrected signals in each data acquisition, event reconstruction processes are essential because the HXI imager consists of a stacked double-sided detector. In double sided strip detectors, data acquisition for one X-ray photon event usually consists of at least two signals, one each from both sides, and these signals often split into 2 adjacent strips. Moreover, some signals are detected in multiple layers due to Cd/Te fluorescence lines or Compton scattering. To identify these multi-signal events, all the signals exceeding the digital thresholds (DTHR) are read out simultaneously from all ASICs in all layers even when only one strip in one layer generates the trigger signal. In the standard HXI analysis, the events are reconstructed as shown in Fig.~\ref{fig:reconstdia}: the gain-corrected EPI in each side are obtained at first, then they are combined within one layer, and then combined with information from other layers to finally reconstruct a photon event.

In the first step, all the signals below analysis thresholds are discarded. The analysis thresholds are larger than DTHR for almost all of the strips, and are set to individual strips, to be four times (DSSDs) or six times (CdTe-DSDs) the standard deviation of pedestal peaks, which correspond to events with zero energy. Thus, the analysis threshold of the DSSDs is much lower than that of the CdTe-DSDs in order to lower the HXI energy range as far as possible. Under these settings, pedestals from all 126 active strips are below the analysis threshold with probabilities of 99.6\% for DSSDs and more than 99.9999\% for CdTe-DSDs. It means that the noise contaminates with probabilities of 0.4\% for DSSD and $<10^{-4}$\% for CdTe-DSD. Mean values of the analysis thresholds of all the strips in the bottom sides (n-sides) of the top-layer DSSD are 3.56~keV for HXI1 and 3.66~keV for HXI2, and they typically distributes from 3~keV to 4~keV. These values determine lower limits of  the energy range of the HXI because noise levels of the bottom sides are worse than those of the top sides in the DSSDs. Thus, the HXI can observe 5~keV in almost all the strips, while there are two strips with analysis thresholds exceeding 5~keV in HXI2. After applying the analysis thresholds, only signals from single strip or two adjacent strips are accepted, and signals from more than two strips and those from non-adjacent two strips are discarded. These events are less than 2--3\% of all the events in the ground data using radioisotopes.

\begin{figure}[tbp]
\begin{center}
\includegraphics[width=\hsize]{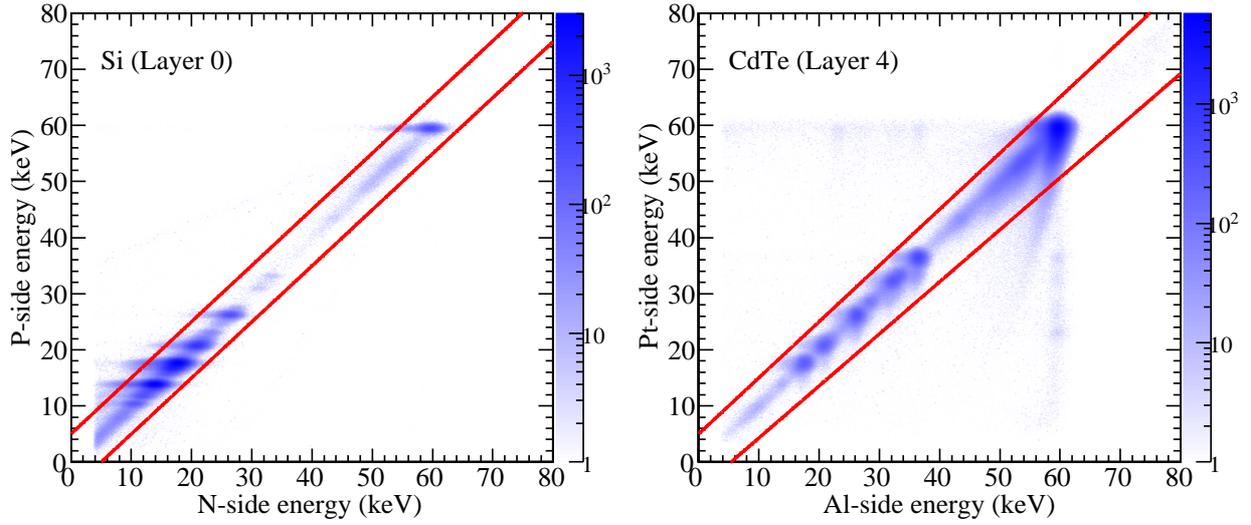}
\caption{Criteria for the consistency check between EPI values from both sides (red lines). Color maps show two-dimensional histograms of the EPIs of HXI2 obtained from the ground calibrations irradiated by X-rays from $^{241}$Am.}
\label{fig:pn}
\end{center}
\end{figure}

Since the HXI is composed of double-sided detectors, signals from top and bottom sides of the layer must be combined to obtain 2 dimensional positional information. EPI from top side in the DSSDs and that from bottom side in the CdTe-DSD have better energy resolutions, and thus are assigned as the EPI value of each layer. Position is simply determined by using an intersection point of strips in both sides. When two adjacent strips have signals, the strip with larger pulse height is assigned.

When combining the signals from both sides, consistency between EPI values from both sides is checked using a condition shown in red lines in Fig.~\ref{fig:pn}. Non-X-ray signals by the instrumental noise or a certain cosmic particles can be rejected by this consistency check. The condition for this check is that the pulse heights (EPI) from both sides match within 5$\sigma$ of the energy resolution. Specifically, it is written as 
\begin{eqnarray}
p_0{\rm EPI}_{\rm bot}-5\sigma_{\rm bot}\leq {\rm EPI}_{\rm top}\leq {\rm EPI}_{\rm bot}+5\sigma_{\rm bot},\label{eq:enecut}
\end{eqnarray}
where $\sigma_{\rm bot}=\sqrt{{p_1}^2+p_2{\rm EPI}_{\rm bot}}$ is a value to represent the energy resolution of the bottom side. The energy resolution is composed of the energy-independent noise component $p_1$ and the Fano noise $\sqrt{p_2{\rm EPI}_{\rm bot}}$. Since the noise level of the bottom side is typically 1.0~keV, we set $p_1=1.0$~keV. The parameter $p_2$ for the Fano noise is a product of the Fano factor $F$ and the electron-hole pair production energy $\epsilon$. By assuming $F=0.1$\cite{Bale1999,Lepy2000} for both Si and CdTe, the second parameters are calculated as $p_2=F\epsilon=0.00036$~keV (Si), $0.00044$~keV (CdTe), where the pair production energies of $\epsilon=3.6$~eV (Si), $4.4$~eV (CdTe)\cite{Takahashi2001} are used. In addition to these parameters, the low mobility of holes in CdTe is taken into account as a parameter $p_0$ by assuming that every 7.1\% of charges are lost during the drift toward the Pt-side strips from the incident position. Thus, $p_0=0.929$ for CdTe-DSD and $p_0=1.0$ (complete charge collection) for DSSD is assumed. In DSSDs, sub-peak events due to the nonuniform electric field\cite{Takeda2007}, which is described in section~\ref{sect:resp}, are also discarded by this process. In the on-ground calibration experiment using $^{241}$Am radioisotope, 2--3\% of total events in DSSDs, and $\sim1$\% in CdTe-DSDs are discarded. 

After finishing the event reconstruction processes in one layer, hits in 5 layers are reconstructed as a photon event. In this process, single-hit events detected in a single layer and double-hit events at the combination of one CdTe-DSD and one DSSD, with an energy of DSSD consistent with a fluorescence line of Cd or Te are accepted. Otherwise, NULL values are recorded in PI, hit positions in the final event list and hence discarded in the following processes. In terms of physical processes, this algorithm accepts photo absorption events and fluorescence escape events, where K-shell fluorescence photons of Cd or Te escaped from CdTe-DSDs are photoabsorbed in DSSDs. Compton scattered events are ignored in current implementation because a fraction of such events composed of Si-Si double hits or Si-CdTe non-fluorescence double hits in total events are less than $\sim1$\% in the ground data obtained with $^{241}$Am and $^{133}$Ba radioisotopes.

\begin{figure}[tbp]
\begin{center}
\includegraphics[width=0.8\hsize]{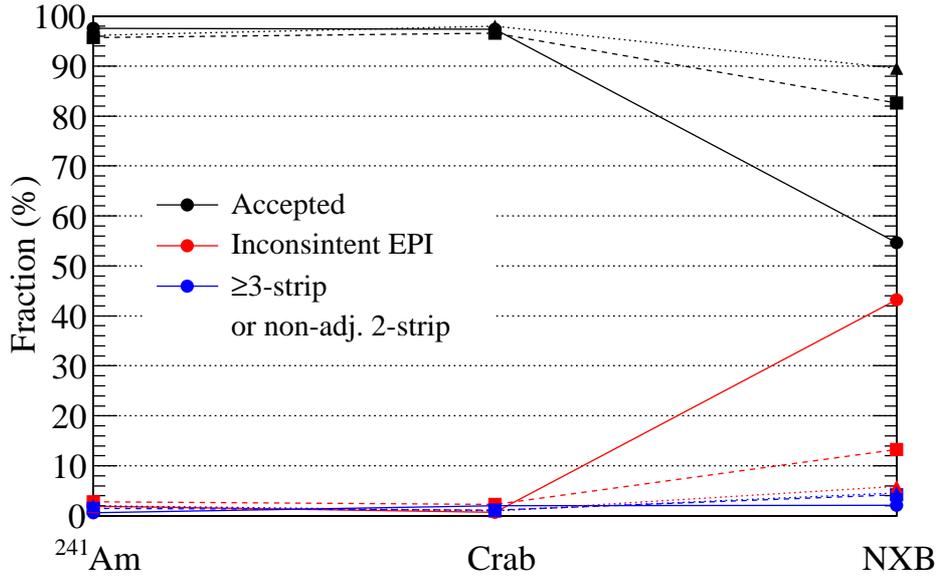}
\caption{Fractions of each event type in the reconstruction for top-layer DSSD (Layer 0), the other DSSDs (Layer 1--3) and CdTe-DSD (Layer 4) among the non-zero signal events. HXI2 data of ground calibration, Crab observation and NXB are used. Filled circles with solid lines, squares with dashed lines and triangles with dotted lines denote the Layer 0, Layer 1--3 and Layer 4, respectively.}
\label{fig:reconst}
\end{center}
\end{figure}

The event reconstruction algorithm described above must be tested with in-flight data because it was determined based on the ground data analysis. For the purpose of investigating whether this algorithm properly rejects the background data without excluding much of the real X-ray signals, fractions of accepted events (black) and discarded events (red and blue) of the ground calibration data (trigger rate $\simeq630$~Hz), Crab data (trigger rate $\simeq610$~Hz) and the NXB data (trigger rate $\simeq35$~Hz) are shown in Fig.~\ref{fig:reconst}. In this figure, the denominator of the fractions is the number of events in which at least one signal exceeds the analysis thresholds. The non-signal events account for $5.341 \pm 0.007$\% of ground calibration data, $25.27 \pm 0.02$\% of Crab data and $97.680 \pm 0.008$\% of the NXB data. These non-signal events are thought to originate from noise triggers and soft photons below the analysis thresholds because trigger thresholds are set to be as low as possible within a range where the dead time fraction due to noise triggers does not affect the scientific observations. The reason it is very high in the NXB data is simply because its trigger rate by external photon and particle background is much lower than those by the instrumental noise (typically a few Hz).

From Fig.~\ref{fig:reconst}, the fraction of discarded events over all non-zero signal events is much larger in the NXB data than the ground calibration and Crab data. All inconsistent EPI, non-adjacent 2-strip and $\geq3$-strip are contained in the NXB data, indicating non-X-ray origins of the signal. On the other hand, more than $\simeq95$\% of events are accepted in ground calibration and Crab data, which are presumably dominated by X-ray signals. Thus, in other words, the multi-layer nature of the imager and the event reconstruction procedure using their information are effective in reducing the background. 

\subsection{Screening}
Besides the event reconstruction process, bad events (e.g., veto events) and bad time intervals (e.g., SAA passages), which are presumably dominated by the instrumental noise or the NXB, are excluded both in the onboard software and the ground analysis software. Basically, the in-flight screening is less stringent than the on-ground screening to flexibly change the screening conditions after observations, on the ground.

The in-flight screening of the HXI is performed in ASICs and HXI Digital Electronics (HXI-DE). In ASICs, only signals exceeding a pre-defined digital threshold, DTHR, are read out. DTHR is adjustable for each ASIC independently, and is also independent from the trigger threshold. On the day before the Crab observation, it was raised up to similar level to the ground-software analysis thresholds for reducing the data size. Read-out data from the ASICs are reduced by further screening in HXI-DE. It assigns ``CATEGORY'' of High, Middle  and Low to each event. Assignment of the CATEGORY is performed using time interval from the previous trigger, number of signals above digital thresholds, ADC values, flags of active shield coincidence (fast BGO and HITPAT BGO), trigger pattern, and the other flags from the ASIC. This CATEGORY determines priorities to record the event to the data recorder (DR) of the satellite. Since the capacity of the DR is limited, most of the data in Middle and Low categories are not downloaded to the ground except for those obtained within the interval the satellite is in direct contact from the operation site at Uchinoura in Japan.

\begin{table*}[tbp]
\caption{Criteria for each CATEGORY}
\begin{center}
\begin{tabular}{lllll}
\hline\hline
CATEGORY & criteria \\
\hline\hline
High & Calibration source, pseudo trigger, test pulse or forced trigger\\
& Not assigned to Middle or Low\\
\hline
Middle & Flagged as HITPAT BGO \& not flagged as fast BGO\\
\hline
Low & Flagged as fast BGO\\
& Number of signals $>31$\\
& Any of signal has ADC $=1023$\\
\hline\hline
\end{tabular}
\end{center}
\label{tab:category}
\end{table*}%

In all observations of the HXI, the screening criteria as listed in Table~\ref{tab:category} is applied in HXI-DE. If an event has a flag for the calibration source signal, pseudo trigger, test pulse or forced trigger, {\tt CATEGORY=High} is assigned automatically. Then, {\tt CATEGORY=Low} is assigned to an event if it has a flag of fast BGO, number of signals exceeds 31 which cannot happen with X-ray photons, or any of signals has ADC value of 1023 which is the upper limit of ADC. In the remaining events, if a signal has a flag of HITPAT BGO, {\tt CATEGORY=Middle} is assigned. While the fast BGO signal is a veto signal capable to stop ADC conversion if needed, the HITPAT BGO signal is slow but lower threshold veto signal to be used for further background rejection in the on-ground data screening (see Ohno~et~al.\cite{Ohno2016} for more details). Finally, {\tt CATEGORY=High} is assigned to all the remaining events. In this screening, only events which are almost certainly the background signals are classified to Middle or Low categories.

The in-flight screening criteria by HXI-DE should also be tested with the Crab and NXB data. In the Crab observation, fractions of High, Middle and Low categories are 93.8\%, 1.7\% and 4.5\%, respectively. Here, these fractions are calculated from the number of events categorized as High/Middle/Low, which are recorded in the housekeeping (HK) data even when the event data is not downloaded. On the other hand, in NXB observations, those are 53.7\%, 1.5\% and 44.8\%, respectively. The fraction of Middle category is very small as expected. Since the Low category contains only background-like events, it is as expected that more events are classified to Low category in the NXB observation than the Crab observation. 

More detailed screening is applied to the HXI data in the on-ground pipeline analysis. In the standard screening criteria for the scientific observations, time intervals at or around the SAA passages and those in Earth occultations are excluded. After the launch of the satellite, a new screening condition {\tt SAA2\_HXI==0} is added to the standard screening to reduce the background in top-layer DSSDs, which is described in section~\ref{sect:nxb}.

\begin{figure}[tbp]
\begin{center}
\includegraphics[width=0.8\hsize]{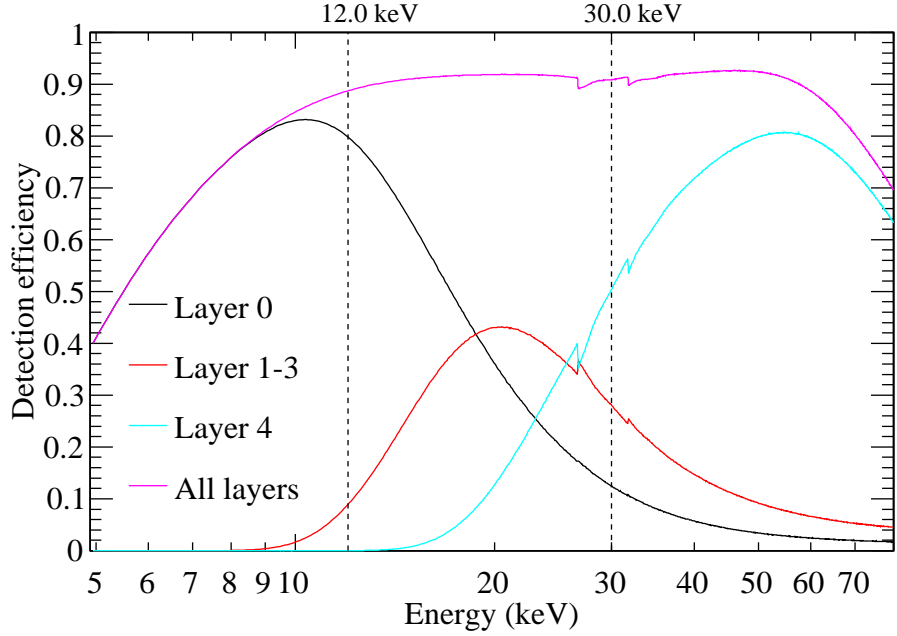}
\caption{Detection efficiency of top-layer DSSD (Layer 0), the other DSSDs (Layer 1--3) and CdTe-DSD (Layer 4). By the energy-dependent layer selection, only events with {\tt PI<300} for Layer 0, {\tt 120$\le$PI} for Layer 1--3, and {\tt 300$\le$PI} for Layer 4 are accepted. The PIs of 120 and 300 correspond to energies of $12$~keV and $30$~keV shown by dotted lines.}
\label{fig:eff}
\end{center}
\end{figure}

Since the HXI imager is a multi-layer detector, an energy-dependent layer selection is applied for maximizing its sensitivity (signal-to-noise ratio). As shown in Fig.~\ref{fig:eff}, the detection efficiency strongly depends on the layer as well as the incident X-ray energies. For example, the top-layer DSSD (Layer 0) is capable of detecting only low energy photons, typically below 30~keV, whereas the CdTe-DSD (Layer 4) covers energies above $\sim20$--$30$~keV. Although the best detection efficiency can be achieved by using all the layers in all energy bands, the background level would inevitably be maximized. Thus, in order to maximize the sensitivity of the HXI, the energy-dependent layer selection is necessary. As explained in detail in section~\ref{sect:nxb}, the NXB of the top-layer DSSD (Layer 0) is different from DSSDs in the other layers due to the electron background, while those in the middle layers (Layer 1--3) are very similar to each other. Thus, the HXI imager is separated into three groups, Layer 0, Layer 1--3 and Layer 4, and their sensitivities are estimated by using the in-orbit NXB spectra and the effective area. To optimize these sensitivities, only events satisfying conditions of {\tt PI<300} for top-layer DSSD (Layer 0), {\tt 120$\le$PI} for the DSSDs in lower layers (Layer 1--3), and {\tt 300$\le$PI} for the CdTe-DSD (Layer 4) are accepted. These correspond to $<30$~keV for the top-layer DSSD, 12--80~keV for the DSSDs in lower layers and $>30$~keV for the CdTe-DSD.

\subsection{Dead Time Correction}
The dead-time correction of the spectra and light curves of the HXI is performed by utilizing pseudo events\cite{Kokubun2006}. The pseudo events are the events triggered by pseudo trigger which is generated with a random time interval in the on-board FPGA in HXI-AE (HXI Analog Electronics). Frequency of the pseudo trigger is set to be 2~Hz by default. Since the pseudo events are treated in the same manner as normal events triggered by the ASICs, the number of pseudo events passing through the data screening divided by the number of input pseudo triggers is a good estimate for the livetime fraction. The process of the dead-time correction is implemented in a dedicated tool {\sc hxisgddtime}.

The average dead time per event in the HXI is $\simeq370$~$\mu$s based on the Crab observation data. It is dominated by reset wait time (250~$\mu$s), which is a wait time to return to the state for waiting the next trigger after the previous data acquisition in order to avoid triggering the noise induced by the AD conversion of the previous event. The other components of the dead time are the AD conversion ($\simeq20\textrm{--}200$~$\mu$s, depending on the pulse height), data transfer from ASICs to the FPGA ($\simeq20$--40~$\mu$s, depending on the number of signals above the DTHR for one trigger) and so on. In addition to the dead time accompanying each event, events in accidental coincidence with BGO active shields generate the dead time.

In the Crab observation, dead time fractions estimated by the pseudo events are 23.4\% and 26.2\% for HXI1 and HXI2, respectively. Their uncertainties are 1--2 percentage point due to the inherent statistical uncertainties of the number of random pseudo events. Then we verified this number with independent estimation. Since the trigger rates of HXI1 and HXI2 are $572.44\pm0.26$~Hz and $613.06\pm0.27$~Hz, respectively, fractions of the dead time depending on the event rate are calculated as $f_{\rm dt}=\tau f=21.0$\% and 22.9\%. Here, $\tau$ is the mean dead time per event (367~$\mu$s for HXI1 and 374~$\mu$s for HXI2), and $f$ is the trigger rate. In addition to this, the fraction of the accidental coincidence with the BGO is estimated by using the number of events classified to Low or Middle. By assuming that rates of Low or Middle events in the Earth occultations do not include the accidental coincidence events, but they are included in the Low/Middle events in the Crab observation, rates of the accidental coincidence events in the Crab observation is estimated by subtracting the Low/Middle rates in the Earth occultations (15.2~Hz for HXI1 and 15.5~Hz for HXI2) from those in the Crab observations (38.5~Hz for HXI1 and 38.1~Hz for HXI2). The accidental coincidence rates are calculated to be 23.3~Hz for HXI1 and 22.6~Hz for HXI2, meaning that 4.05\% and 3.69\% of the trigger rate are discarded. Thus, by considering the accidental coincidence events, the dead time fraction is estimated to be 24.2\% for HXI1 and 25.7\% for HXI2, which match to those estimated by the pseudo events within the statistical uncertainties.

\section{Energy Response Matrix} \label{sect:resp}
\subsection{Simulations of Detector Devices}
The response matrix of the HXI is constructed by running Monte Carlo simulations since Compton scattering and secondary emissions are non-negligible in hard X-ray bands. The simulation is composed of two steps: Monte-Carlo simulations for calculating the interactions of X-ray photons with detectors and passive materials in the HXI system, and calculations of charge transportation in the semiconductor detectors. This simulation code is based on an integrated response generator ``ComptonSoft'' \cite{Odaka2010}, which is available at a web-based repository on GitHub\footnote{\linkable{https://github.com/odakahirokazu/ComptonSoft}}.

\begin{figure}[tbp]
\begin{center}
\includegraphics[width=0.8\hsize]{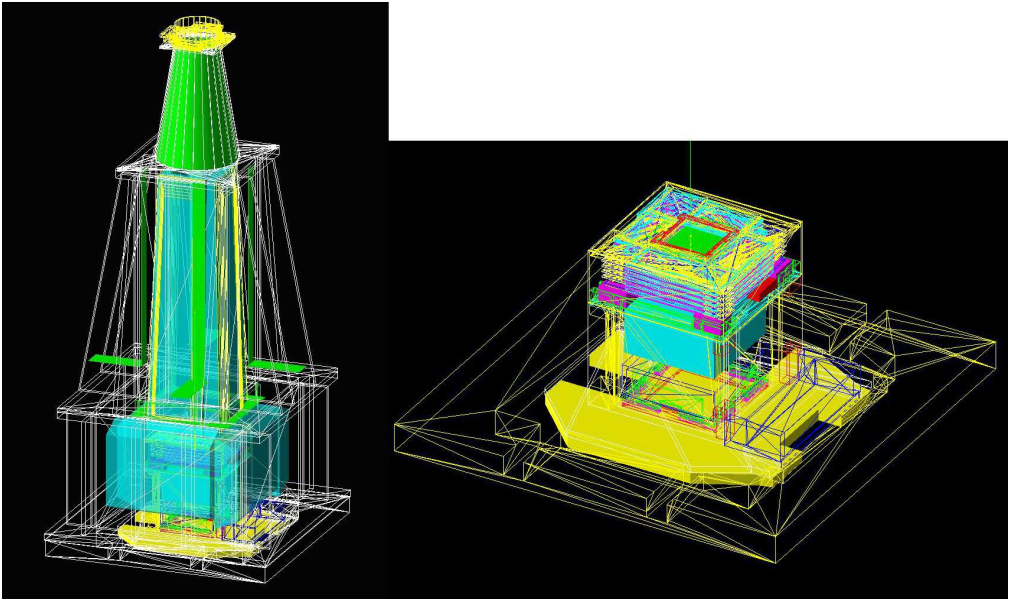}
\caption{Geant4 mass model of the HXI (left) and that of the imager module (right) for the detector response simulations.}
\label{fig:massmodel}
\end{center}
\end{figure}

The Monte Carlo simulation part is based on the Geant4 toolkit library\cite{Agostinelli2003,Allison2006}, which is widely used for the particle tracking in high-energy physics. Fig.~\ref{fig:massmodel} shows a detailed Geant4 mass model of the HXI implemented in the detector response simulations. Since all the materials in the HXI system affects the detector response, most of the passive materials as well as the imager module and BGO active shields are included in the mass model. The Monte Carlo simulations are performed for each energy bin of the response matrix with monochromatic photons at the central energy of the bin. The incident photons for the simulations are generated in a horizontal plane with the detector size ($32\times32$~mm$^{2}$) located above the HXI-S entrance window, with an initial direction to the detector along the optical axis. 

After the Monte Carlo simulations, the energy deposits obtained in the simulations are spread by considering the thermal diffusion. In this simulation, it is spread by the 2-dimensional Gaussian with $\sigma=f_{\rm diff}\sqrt{2k_{\rm B}T\mu t/e}$, where $k_{\rm B}$ is the Boltzmann constant, $T$ is the temperature of the detector and $t$ is the drift time from carrier creation to arrival to the electrodes. An additional factor $f_{\rm diff}$ is introduced to reproduce experimental results. Since it only considers the thermal diffusion, this factor represents an effect by the Coulomb repulsion\cite{Benoit2009,Kitaguchi2011}.

By using the spread energy deposits, the induced charge on each read-out strip are calculated with the simulation of the charge transportation. The induced charge $Q$ is calculated utilizing a weighting potential $\phi_{\rm w}$ derived from the Shockley-Ramo theorem\cite{He2001} as
\begin{eqnarray}
Q=-\int q(\bm{x})\bm{\nabla}\phi_{\rm w}(\bm{x})\cdot d\bm{x},
\end{eqnarray}
where $q(\bm{x})$ is a charge inside the detector and $\int d\bm{x}$ is an integration along the trajectory of the charge $q(\bm{x})$ to the read-out electrode. By considering the finite lifetime $\tau$ and mobility $\mu$ of carriers and assuming an uniform electric field $E$ in the detectors, the induced charge is written as
\begin{eqnarray}
Q=-q_0\int_{z_{\rm i}}^{z_{\rm f}}\exp\left(-\frac{z-z_{\rm i}}{\mu\tau E}\right)\frac{\partial\phi_{\rm w}(x,z)}{\partial z}dz,\label{eq:charge}
\end{eqnarray}
where $z_{\rm i}$ and $z_{\rm f}$ are the initial and final positions of the charge and $q_0$ is the initial charge. The strip electrodes are lined up in $x$ direction and $z$-axis points down into the detector. The induced charge is calculated by multiplying the energy deposits and the charge collection efficiency defined as a sum of $Q/q_0$ for holes and electrons, which depends on the interaction position.

The weighting potential is a solution of Laplace's equation with the boundary condition
of $\phi_{\rm w}=1$ at the readout electrode and $\phi_{\rm w}=0$ at all the other electrodes.
That for the strip detectors is calculated as
\begin{eqnarray}
\phi_{\rm w}(x,z)&=&\sum_{m=1}^{\infty}A_{m}\sin\left(\alpha_mx\right)\sinh\left(\alpha_mz\right)\label{eq:trap}\\
A_m&=&\frac{2}{m\pi\sinh(\alpha_m L)}f_m\\
f_m&=&\cos\left[\frac{\alpha_m(a-U)}{2}\right]-\cos\left[\frac{\alpha_m(a+U)}{2}\right],
\end{eqnarray}
where $\alpha_m=m\pi/a$, $a$ is the detector size, $U$ is the strip pitch and $L$ is the thickness of the detector. In the case of the HXI CdTe-DSDs, $a=32$~mm, $U=250$~$\mu$m and $L=750$~$\mu$m. By combining Eq.~\ref{eq:charge} with Eq.~\ref{eq:trap}, the charge collection efficiency of CdTe-DSD is calculated. 

The response of the HXI DSSDs is more affected by charge loss due to the complicated electric field structure inside the detector rather than the charge trapping due to finite lifetime of the carriers described in Eq.~\ref{eq:charge}. Since a mobility-lifetime product ($\mu\tau$) of carriers in Si is 2--3 orders of magnitude larger than that of CdTe, it is assumed to be infinite in this simulation, which is equivalent to the charge correction efficiency of unity. On the other hand, a significant fraction of the charge is lost by local minimum of the electric potential due to a positive fixed charge on the Si--SiO$_2$ surface at gaps between strip electrodes\cite{Takeda2007,Miyake2016}. It makes a sub-peak at $\simeq1/2$ of the incident X-ray energy in the top side of DSSD, negative peak in the adjacent strip, and no signals in the bottom side of DSSD. Thus, we refer to these events as sub-peak events. The sub-peak events are unusable for the event reconstruction because they do not have any information of the position in the bottom side. Specifically, they are discarded in the consistency check between EPI values from both sides (Eq.~\ref{eq:enecut} and Fig.~\ref{fig:pn}). Since the potential local minimum is located on surface of the top side of the detectors, this effect reduces the detection efficiency at energies below $\sim10$~keV. This effect is simply implemented as rectangular dead regions located at the strip gaps on the surface of the DSSD in our simulation. All induced charges corresponding to energy deposits in these dead regions are set to be zero.

\subsection{Simulation Parameters}
All the parameters of the detector simulations for constructing the response matrix are listed in Tab.~\ref{tab:params}. The bias voltage, the mobility-lifetime products $\mu\tau$ of holes and electrons, the diffusion factor, and the noise level of each strip are required for both CdTe-DSD and DSSDs, and one additional parameter for sub-peak events is required for DSSDs. In addition to these, trigger efficiency, which reduces the detection efficiency at the lower energy end due to the energy resolution in the shaper for the trigger generation, and absorption by SiO$_2$ layers on surface of the DSSDs are multiplied to the energy response.

\begin{table}[tbp]
\caption{Simulation parameters for the HXI detector response}
\begin{center}
\begin{tabular}{lcccccc}
\hline
Parameters* & \multicolumn{6}{c}{Values}\\
\hline
& \multicolumn{3}{c}{HXI1} & \multicolumn{3}{c}{HXI2}\\
Layer & 0 & 1--3 & 4 & 0 & 1--3 & 4\\
\hline
Bias [V] & 250 & 250 & 250  & 250 & 250 & 350\\
$(\mu\tau)_{\rm h}$ [cm$^2$ V$^{-1}$]& $\infty$ & $\infty$ & $1.54\times10^{-4}$ & $\infty$ & $\infty$ & $1.54\times10^{-4}$\\
$(\mu\tau)_{\rm e}$ [cm$^2$ V$^{-1}$]& $\infty$ & $\infty$ & $1.41\times10^{-3}$ & $\infty$ & $\infty$ & $1.41\times10^{-3}$\\
$d_\textrm{sub-peak}$ [$\mu$m]& 25 & 25 & ---  & 25 & 25 & ---\\
$f_{\rm diff.}$ & 2 & 2 & 2 & 2 & 2 & 2\\
$d_{\rm SiO_2}$ [$\mu$m]& 5 & 4 & --- & 3 & 4 & ---\\
$E_{\rm trig.}$ [keV]& 2.97 & --- & --- & 3.05 & --- & ---\\
$\sigma_{\rm trig.}$ [keV]& 0.92 & --- & ---  & 0.70 & --- & --- \\
Top-side noise [keV]& 0.86 & 0.90 & 1.71 & 0.89 & 0.90 & 1.84\\
Bottom-side noise [keV]& 2.49 & 3.04 & 1.75 & 2.57 & 2.92 & 1.80\\
\hline
\hline
\end{tabular}
\end{center}
\begin{flushleft}
\footnotesize
* $(\mu\tau)_{\rm h}$ and $(\mu\tau)_{\rm e}$ are the mobility-lifetime products of holes and electrons, $d_\textrm{sub-peak}$ is a thickness of the dead region where the sub-peak events are generated, $f_{\rm diff.}$ is the diffusion factor for spreading carrier clouds, $d_{\rm SiO_2}$ is the thickness of the ${\rm SiO_2}$ layer on surface of the DSSDs, $E_{\rm trig.}$ and $\sigma_{\rm trig.}$ are the mean energy and $\sigma$ of an error function for describing the trigger efficiency.
\end{flushleft}
\label{tab:params}
\end{table}%

The parameters used in the HXI response simulations are determined based on the ground calibration data of flight model and engineering model of the HXI. The mobility-lifetime product $\mu\tau$ of CdTe-DSD is determined by fitting spectra of the single layer experiment of the engineering model detectors. The diffusion factor $f_{\rm diff.}$ is set to be 2 as it reproduces the engineering model data. Noise parameter of each channel is estimated from the line widths of an X-ray line at 59.5~keV from $^{241}$Am obtained in the ground calibration tests of the flight model detectors. Although only mean values of the noise levels from all the strips are listed in Tab.~\ref{tab:params}, the noise level is assigned strip by strip in the simulation.

Detection efficiency at low energies of DSSDs is affected by three effects: sub-peak events, trigger efficiency, and absorption by inactive layer. Size of the dead region due to the sub-peak events is assumed to be a rectangle with a width of 120 $\mu$m, which is same as the width of gap between the strip electrodes of the DSSDs. By performing experiments using a single layer of the DSSD engineering model, thickness of the dead region $d_\textrm{sub-peak}$ is estimated to be 25~$\mu$m from the energy dependence of the sub-peak fraction\cite{Miyake2016}. Trigger efficiency of the top-layer DSSD is assumed to follow the error function, and its mean energy and $\sigma$ are estimated by measuring the detected count rate for 5.9~keV line from $^{55}$Fe and its sub-peak at 3.2~keV as a function of the trigger threshold in the ASIC. The trigger generation and sample/hold (and then the ADC) are performed in different analog shaping chains in the ASIC with different shaping times of $0.6$~$\mu$s for the former part and $\simeq3$~$\mu$s for the latter. The noise level of the trigger $\sigma_{\rm trig.}$ is usually worse than that of the EPI values corresponding to the spectral resolution. In the other layers, trigger efficiency is not considered because the trigger threshold $E_{\rm trig.}$ is much lower than the analysis threshold applied in the pipeline process. The thickness of inactive layer including SiO$_2$ layers and Al electrodes on surface of the DSSDs is estimated to be $\simeq4$~$\mu$m. Since difference between photoabsorption cross-sections with SiO$_2$ and Al is negligible, the inactive layer is treated as a SiO$_2$ layer with a thickness of $d_{\rm SiO_2}$. Thus, this value is set to the detectors in Layer 1--3 as listed in the table.

In addition to the relatively rough estimates of the sub-peak fraction, trigger efficiency and the SiO$_2$ thickness, more detailed tuning of these parameters is needed to reproduce the in-flight data, especially around the lower energy end, in which the photon statistics is the highest in many cases. Among these three parameters, SiO$_2$ thickness is chosen as a free parameter for adjusting. First, 5--12~keV spectra of G21.5$-$0.9 observed by the SXI and HXI are simultaneously fitted, and the SiO$_2$ thicknesses of HXI1 and HXI2 are constrained to be 3.7--5.6~$\mu$m and 2.5--4.4~$\mu$m as 90\% confidence intervals. Then, in order to constrain the SiO$_2$ thickness more tightly, the 5--40~keV Crab spectra of the HXI is fitted by using response matrices with SiO$_2$ thickness of 4.0, 4.5, 5.0 and 5.5~$\mu$m for HXI1, and 2.5, 3.0, 3.5 and 4.0~$\mu$m for HXI2. As the result, $d_{\rm SiO_2}=5.0$~$\mu$m for HXI1 and $d_{\rm SiO_2}=3.0$~$\mu$m for HXI2 are found to give the best $\chi^2$ value. Thus, these values are used for constructing the HXI response matrix. This difference between HXI1 and HXI2 changes the detection efficiency at 5~keV by 7\%. Please note that this result does not mean that the actual SiO$_2$ thickness is different between HXI1 and HXI2. The difference would include all of the effect in the lower energies by the sub-peak fraction, trigger efficiency and the SiO$_2$ thickness.

\subsection{Validation with Ground Calibration Data}
\begin{figure}[tbp]
\begin{center}
\includegraphics[width=\hsize]{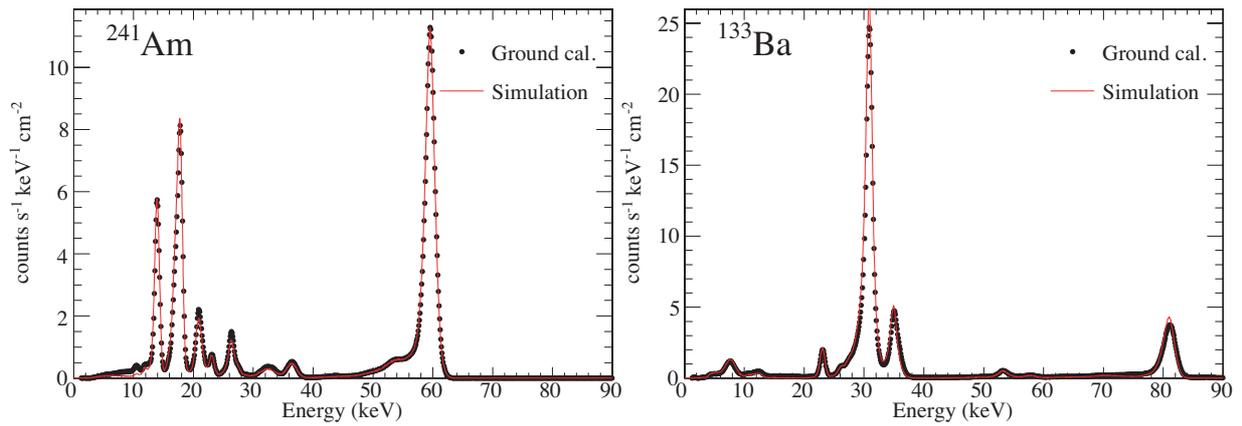}
\caption{HXI2 spectra of X-ray lines from $^{241}$Am and $^{133}$Ba. Black points are the experimental spectra obtained in the Ground calibrations of the HXI flight model, and red lines are the simulated spectra.}
\label{fig:simspec}
\end{center}
\end{figure}

\begin{figure}[tbp]
\begin{center}
\includegraphics[width=\hsize]{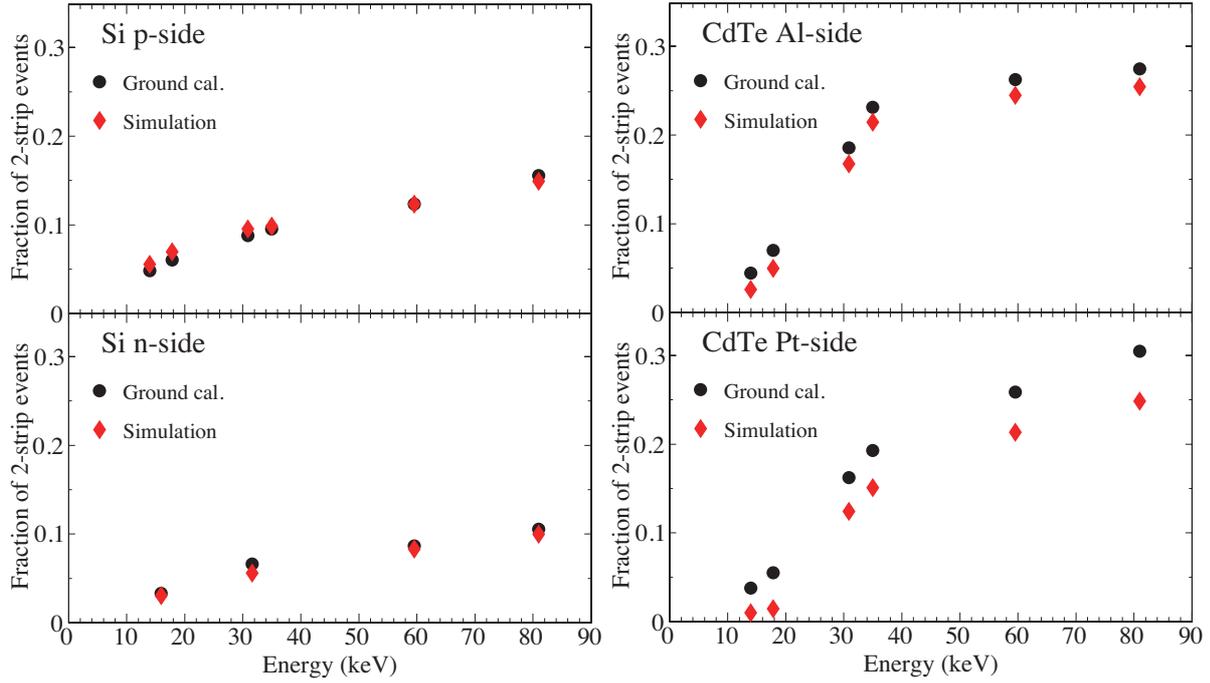}
\caption{Fractions of the 2-strip events as a function of energy in the ground calibration data (black circle) and the simulated data (red diamond).}
\label{fig:2strip}
\end{center}
\end{figure}

For the purpose of testing the response simulation and its parameters, the simulated spectra and their properties are compared with those of the ground calibration data irradiated by X-rays from $^{133}$Ba and $^{241}$Am. The simulations are performed in a geometry with a thermostat chamber and passive materials near the radioisotopes. All the X-ray and Gamma-ray lines above 10~keV with emission probabilities larger than 0.1\% are simulated. The Lund/LBNL Nuclear Data Search\footnote{\linkable{http://nucleardata.nuclear.lu.se/toi/}} is referred for the energies and emission probabilities of all the lines from $^{133}$Ba and 26.3~keV and 59.5~keV from $^{241}$Am, while L\'{e}py et al (2008)\cite{Lepy2008} is referred for the other lines from $^{241}$Am.

The simulated spectra and the experimental spectra from the ground calibrations of the HXI flight model are shown in Fig.~\ref{fig:simspec}. Spectral shapes, including scattered components, tail structures due to the small $\mu\tau$ in CdTe and energy resolutions are well reproduced. Moreover, detection efficiency is consistent within $\simeq10$\%, and at higher energies above $\simeq60$~keV, it matches better than 5\%. Here, we should note that the self-absorption effect in the $^{241}$Am source is considered in this simulation by assuming a $5$-$\mu$m-thick Am as an absorber. Fig.~\ref{fig:2strip} shows fractions of the 2-strip events of the simulations and the experiments. These fractions have to match with each other for reproducing the spectral shape because energy resolutions of the 2-strip spectra are worse than those of the single-strip spectra. In other words, if the simulation overestimates the 2-strip fraction, the energy resolution of simulated spectra would be worse than those of the experiments. As shown in the upper panels, the 2-strip fractions are reproduced within 2\% of the total event number in DSSD p-side and CdTe-DSD Al-side, whose EPI values are used for the spectral analysis. The discrepancy in CdTe-DSD Pt-side has almost no effects on the scientific analysis because the EPI values from CdTe-DSD Pt-side and DSSD n-side are only used for the consistency check between signals in top and bottom sides in the event reconstruction process.

\subsection{Crab Spectra}
Using the HXI response matrix described above, the Crab spectra are analyzed. All the standard processing and screenings (see section~\ref{sect:ana}) are applied to the data, and the spectra are extracted from circular regions with a radius of 4$^\prime$. The background spectra are extracted from the blank sky observations (i.e., None2, IRU check out and RXJ1856.5$-$3754), which contain the cosmic X-ray background as well as the NXB. The size of the extraction regions for the background spectra are the same as those for the Crab spectra. Net exposures of the Crab spectra after the dead time correction are 5.92~ks for HXI1 and 6.14~ks for HXI2, and $2.69\times10^6$ photons are detected by each of HXI1 and HXI2 in an energy range of 5--80~keV. 

\begin{figure}[tbp]
\begin{center}
\includegraphics[width=0.8\hsize]{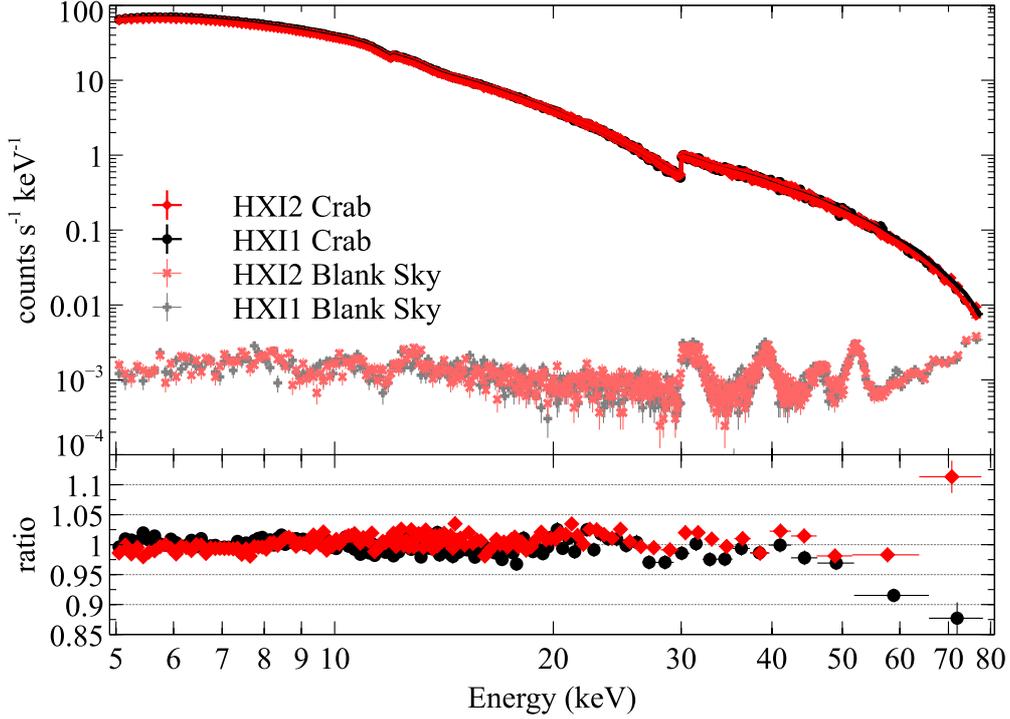}
\caption{Crab and background spectra observed by HXI1 and HXI2 and the ratio between the Crab spectra and an absorbed powerlaw model.}
\label{fig:Crab}
\end{center}
\end{figure}

The Crab spectra observed by the HXI are shown in Fig.~\ref{fig:Crab}. They are fitted with an absorbed powerlaw model {\sc constant*tbabs*powerlaw} convolved with the detector response and the telescope effective area using the spectral analysis software {\sc XSPEC}\cite{Arnaud1996}. As shown in the lower panel in Fig.~\ref{fig:Crab}, the deviations between the Crab spectra and the best-fit model are less than 5\% at energies below $\simeq50$~keV. The telescope effective area is measured with uncertainties less than $\simeq 2$\% on ground\cite{Mori2018}, and confirmed by the Crab observation\cite{Matsumoto2017}. Although both the telescopes and detectors are well calibrated, residuals of $\sim13$\% level is seen above 50~keV. This might be due to calibration uncertainties in the telescope effective area or inappropriate modeling of the detector response.

The best-fit parameters and 90\% confidence errors for the HXI Crab spectra are listed in Tab.~\ref{tab:Crabfit}. A difference of normalizations between HXI1 and HXI2, which is expressed by a constant parameter $0.968$, is consistent with unity considering the uncertainty in the dead-time correction. For example, we also applied the independent dead-time correction, as discussed in section~\ref{sect:ana}, which are estimated to be 24.2\% for HXI1 and 25.7\% for HXI2. If we adopt these values, the powerlaw normalization at 1~keV is $N=10.43\pm0.04$~photons~cm$^{-2}$~s$^{-1}$~keV$^{-1}$ and a constant parameter is $f_{\rm HXI2/HXI1}=0.985\pm0.001$.

Our best-fit values of the powerlaw index $\Gamma=2.107\pm0.002$ and a normalization at 1~keV, $N=10.54\pm0.04$ are consistent with the historical values of $\Gamma=2.10\pm0.03$ and $N=9.7\pm1.0$ proposed by Toor \& Seward (1974)\cite{Toor1974}. Also, the spectral slope is consistent with the values of $\Gamma=2.10\pm0.01$ obtained by {\it Suzaku/HXD} PIN with HXD nominal position\cite{Kokubun2006} and $\Gamma=2.106\pm0.006$ obtained by the large off-axis observations of the Crab by {\it NuSTAR}\cite{Madsen2017}. On the other hand, the normalization is not consistent with either of these observations. Our best-fit normalization is just between $N=11.2\pm0.09$ by the HXD and $N=9.71\pm0.16$ by {\it NuSTAR}. This result does not immediately mean that the HXI effective area is inconsistent with the other observatories because the Crab flux can vary on a yearly timescale\cite{Wilson-Hodge2011}.

\begin{table}[tbp]
\caption{Best-fit parameters and 90\% confidence errors of the Crab spectra}
\begin{center}
\begin{tabular}{lccccc}
\hline
& $N_{\rm H}$ & $\Gamma$ & Normalization@1~keV & Constant & $F_\textrm{3--50~keV}$\\
& [cm$^{-2}$]  & & [photons~cm$^{-2}$~s$^{-1}$~keV$^{-1}$] & & [$10^{-8}$~erg~cm$^{-2}$~s$^{-1}$]\\
\hline
HXI1 & \multirow{2}{*}{$3.0\times10^{21}$*} & \multirow{2}{*}{$2.107\pm0.002$} & \multirow{2}{*}{$10.54\pm0.04$}** & 1.0* & \multirow{2}{*}{$3.647\pm0.004$}\\
HXI2	 & & & & $0.968\pm0.001$\\
\hline
\hline
\end{tabular}
\end{center}
\begin{flushleft}
\footnotesize
* Column density $N_{\rm H}$ and the constant factor for HXI1 are fixed.\\
** The statistical uncertainty of the dead time ($\simeq1\textrm{--}2$\% in 1-$\sigma$ confidence level) is not added. It corresponds to the normalization uncertainty of $\simeq 0.2\textrm{--}0.3$ (90\% confidence).
\end{flushleft}
\label{tab:Crabfit}
\end{table}%

\subsection{Spatial dependence of the detector response}
Besides the spatially integrated detector response used in the Crab spectral analysis, we also verified the reproducibility of positional difference of the detector response. In the standard analysis tool, the response is separated into the detection efficiency and response matrix. The matrix defines the relation between the incident photon energy and the output EPI values, which correspond to the spectral shape. The spatial dependence of the detection efficiency is implemented for pixel by pixel based on the Monte Carlo simulations of the detector response. On the other hand, the response matrix is integrated over all the detector area in order to reduce data size of response database files. Therefore, the spatial dependence of the detector response is taken into account only by the detection efficiency.

\begin{figure}[tbp]
\begin{center}
\includegraphics[width=0.8\hsize]{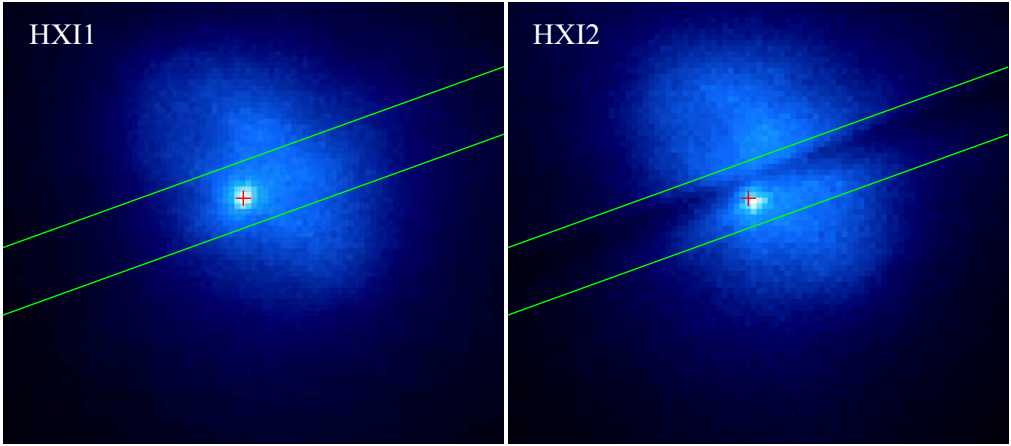}
\caption{Crab images obtained by HXI1 and HXI2. Noisy strips are clearly seen in the HXI2 image. The red cross and green lines indicate the assumed source position for the response simulations and the region from which the spectra are extracted, respectively. Events below 5~keV is also included in this image for emphasizing the noisy strips.}
\label{fig:badstripimage}
\end{center}
\end{figure}

To demonstrate the accuracy of the spatial dependence of the detector response, noisy strips located close to the center of the FoV of HXI2 provide a good example. The analysis thresholds of these noisy strips are 5.58~keV and 6.71~keV, which are much higher than those in the typical strips, 3.66~keV.  Due to the higher analysis thresholds, detection efficiencies at low energy in these strips are significantly smaller than the other strips, resulting in a dark line at the center of the image as shown in Fig.~\ref{fig:badstripimage}.

\begin{figure}[tbp]
\begin{center}
\includegraphics[width=0.8\hsize]{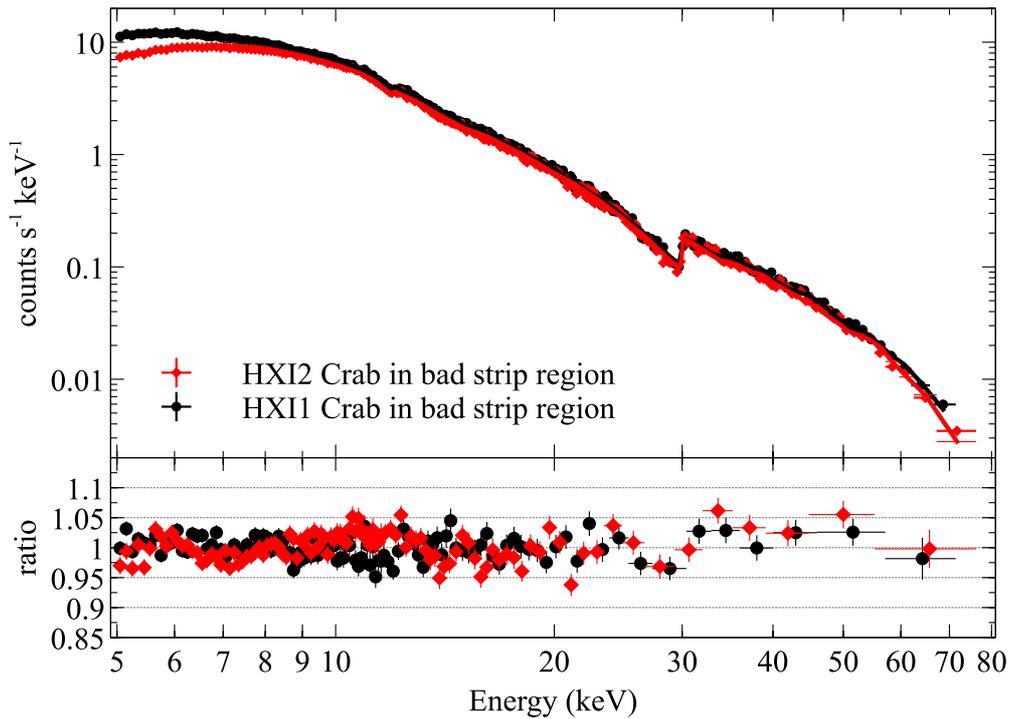}
\caption{Crab spectra extracted from a $20^{\prime\prime}\times540^{\prime\prime}$ rectangular region covering the noisy strips, and the ratio between the spectra and an absorbed broken-powerlaw model.}
\label{fig:badstripspec}
\end{center}
\end{figure}

The spectra extracted from a $20^{\prime\prime}\times540^{\prime\prime}$ rectangular region covering the noisy strips in HXI2 are shown in the upper panel of Fig.~\ref{fig:badstripspec}. There is a clear difference of the low energy spectra between HXI1 and HXI2. The count rate at 5~keV in HXI2 is smaller than that of HXI1 by a factor of 1.5 because of the higher analysis thresholds in the noisy strips in HXI2. As shown in the lower panel in Fig.~\ref{fig:badstripspec}, this large difference between HXI1 and HXI2 is reduced to better than $\pm5$\% level by applying the detector response, which is generated in a standard manner by assuming a point source located at the red cross in Fig.~\ref{fig:badstripimage}. In this analysis, a broken powerlaw model is assumed by following the {\it NuSTAR} observations of the Crab pulsar\cite{Madsen2015a} because the selected spectra is strongly affected by it. The best-fit parameters of powerlaw slopes $\Gamma_1=2.029\pm0.010$ and $\Gamma_2=2.185\pm0.008$, break energy $E_{\rm b}=10.4\pm0.4$~keV,  normalization $N=7.0\pm0.2$~photons~cm$^{-2}$~s$^{-1}$~keV$^{-1}$ and the constant parameter $f_{\rm HXI2/HXI1}=1.067\pm0.004$ are obtained. This result demonstrates the accuracy of the spatial dependence of the HXI detector response.

\section{Non-X-ray Background} \label{sect:nxb}

\subsection{Properties of DSSD Background}
\begin{figure}[tbp]
\begin{center}
\includegraphics[width=0.8\hsize]{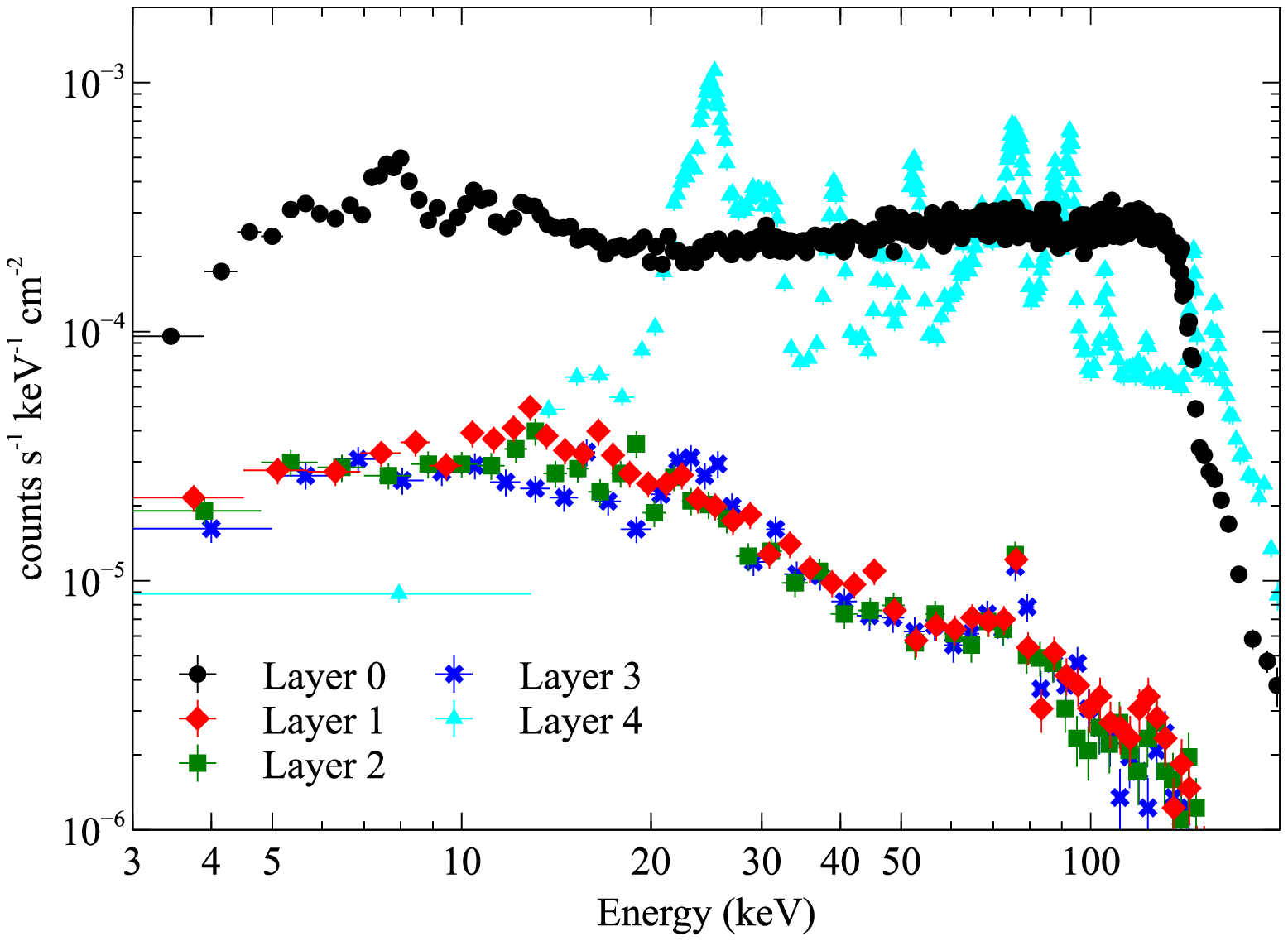}
\caption{HXI2 NXB spectra extracted from Earth occultation data. The spectra are extracted from whole detector area, and scaled with its geometrical area. Please note that a cutoff at $\sim120$~keV in DSSDs is due to the upper limit of the dynamic range of slow shapers in readout ASICs, while that at $\sim150$~keV in CdTe-DSD is due to the upper limit of the ADC in ASICs.}
\label{fig:unscreenedNXB}
\end{center}
\end{figure}
Before the screening of {\tt SAA2\_HXI==0} described in section~\ref{sect:ana}, non-X-ray background (NXB) in top-layer DSSD (Layer 0) of the HXI is dominated by a hard powerlaw component as shown in Fig.~\ref{fig:unscreenedNXB}. In pre-launch estimations, NXB in top-layer DSSD is expected to show similar level with those in the other layers of DSSDs because their background is thought to be mainly caused by albedo neutrons, interacting via elastic scattering\cite{Mizuno2010}. In this sense, the fact that the middle layers (Layer 1--3) of DSSDs show a similar level is as expected.

\begin{figure}[tbp]
\begin{center}
\includegraphics[width=0.8\hsize]{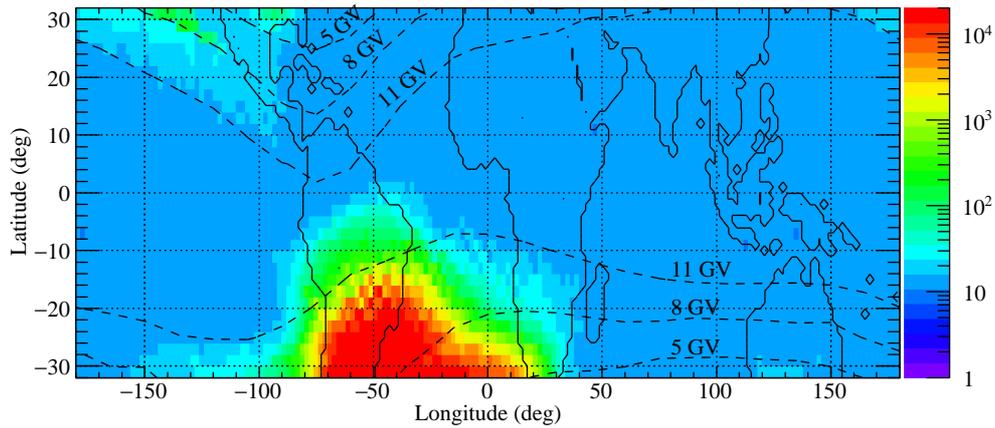}
\caption{Distribution of trigger rate of the HXI2 top-layer DSSD during Earth occultation and blank sky observations. Black dashed lines indicate the geomagnetic cut-off rigidity.}
\label{fig:SAAMap}
\end{center}
\end{figure}

The powerlaw component in top-layer DSSD is due to the low energy albedo electrons for the following two reasons. First, this component extends up to 100~keV as shown in Fig.~\ref{fig:unscreenedNXB}, but lower layers do not show this component strongly. The difference cannot be explained if it is caused by $\sim100$~keV photons. It means that this background component originates from particles with low penetrating power. The second reason is the distribution of the background rate. Fig.~\ref{fig:SAAMap} shows a trigger rate of the HXI2 top-layer DSSD as a function of latitude and longitude of the satellite. It extends larger than the SAA, and has a hot region at above north America. The hot region does not simply depend on the geomagnetic cut-off rigidity because it is not seen in other regions. This distribution is consistent with the electron distribution ($>93$~keV) observed by DEMETER/IDP in orbit\cite{Whittaker2013}.

\begin{figure}[tbp]
\begin{center}
\includegraphics[width=0.8\hsize]{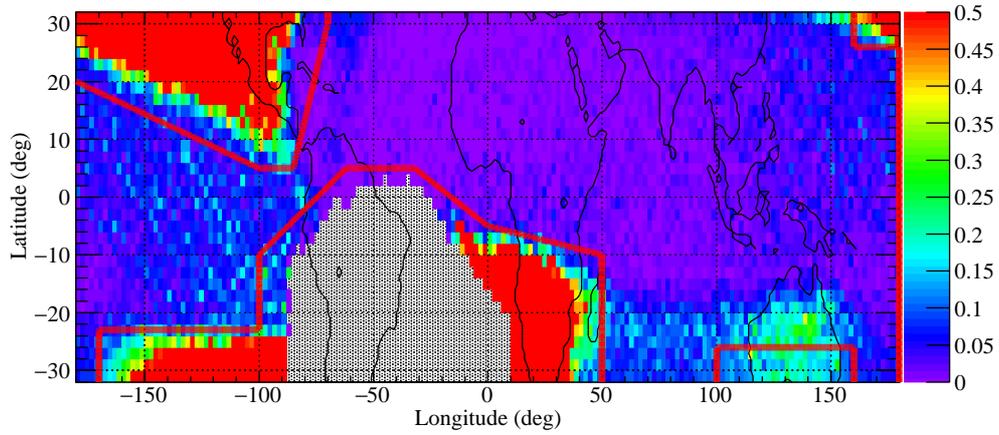}
\caption{Distribution of count rate of the HXI2 top-layer DSSD with $PI>800$ during Earth occultation and blank sky observations. Signals in regions enclosed by red lines are ignored by the satellite position selection. }
\label{fig:SAAMap_evt}
\end{center}
\end{figure}

{\it NuSTAR} might observe this electron background as well. According to a web page on the background filtering\footnote{\linkable{www.nustar.caltech.edu/page/background}}, a ``tentacle''-like region of higher activity is found to be located around $\sim-90^\circ$ longitude and above about $-2^\circ$ latitude. This region is consistent with the distribution of the low energy electrons as shown in Fig.~\ref{fig:SAAMap} and \ref{fig:SAAMap_evt}.

The albedo electrons cannot directly come into the HXI imager because baffles made of Pb/Sn shields are implemented to block the stray light and cosmic X-ray background out of the FoV. Thus, top-layer DSSD probably suffer from the electrons scattering on the HXI-S entrance window (two layers of $30$~$\mu$m-thick poly-Carbonate sheets) or the extensible optical bench. To reduce this background, we should have baffled the entrance window from the low energy albedo electrons.

\begin{figure}[tbp]
\begin{center}
\includegraphics[width=0.8\hsize]{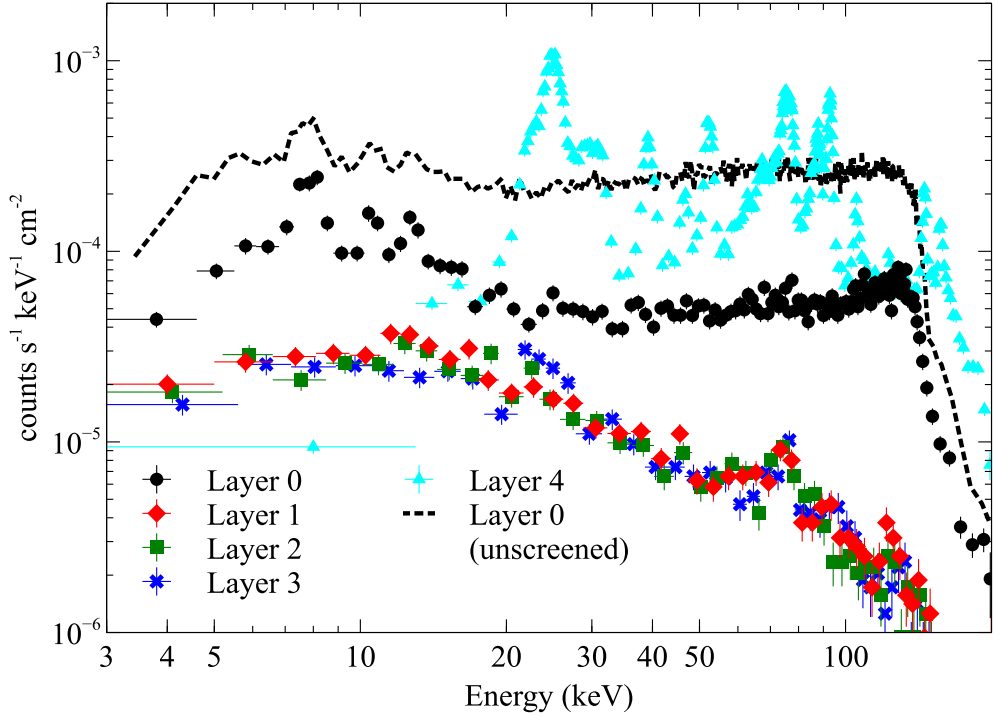}
\caption{HXI2 NXB spectra screened with satellite position. Unscreened NXB spectrum of Layer 0 (top-layer DSSD), which is already shown in Fig.~\ref{fig:unscreenedNXB}, is also overplotted as a comparison.}
\label{fig:screenedNXB}
\end{center}
\end{figure}

Since the electron background strongly depends on the satellite position as shown in Fig.~\ref{fig:SAAMap}, a selection with the satellite position successfully reduces top-layer DSSD background down to $10$--20\% level as shown in Fig.~\ref{fig:screenedNXB}. The definition of the regions discarded in this selection is shown in red lines in Fig.~\ref{fig:SAAMap_evt}. This region is implemented in the standard screening procedure as {\tt SAA2\_HXI}. This selection reduces observation efficiency as well. Area of the selected regions (which excludes the SAA as well) is $\simeq67$\% of the total area where the satellite orbits above, while those of {\it Suzaku} SAA definition is $\simeq88$\%.

\begin{figure}[tbp]
\begin{center}
\includegraphics[width=0.8\hsize]{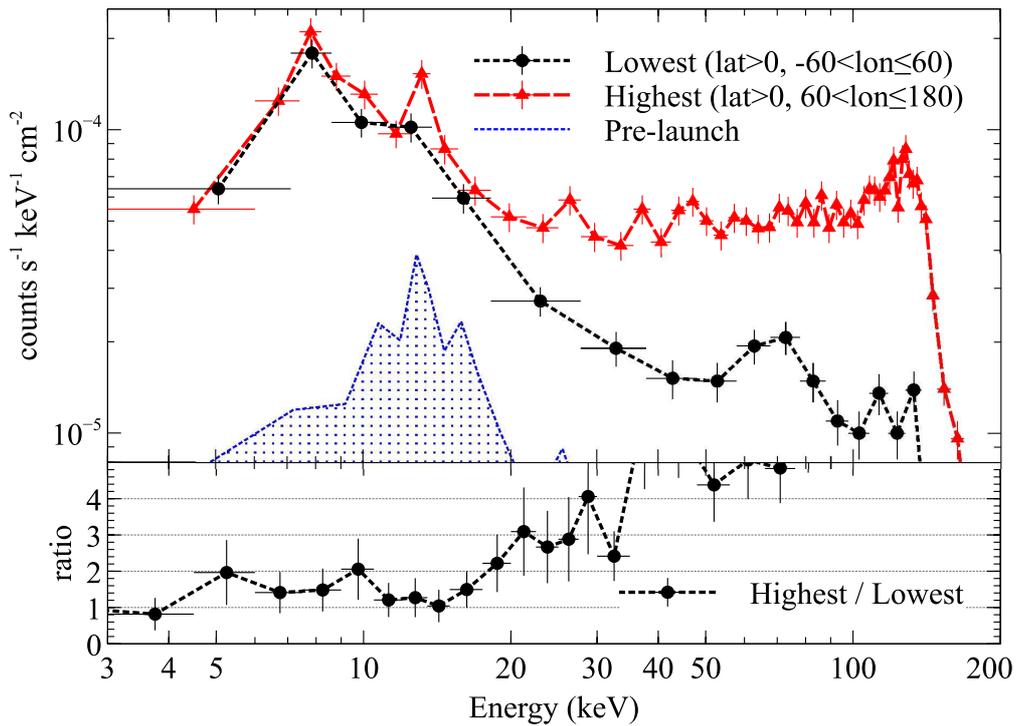}
\caption{HXI2 NXB spectra of top-layer DSSD extracted from high and low background regions, after the final screening.}
\label{fig:highlowE}
\end{center}
\end{figure}

The electron background depends on the satellite position, even after excluding the high background regions using the satellite position selection. Fig.~\ref{fig:SAAMap_evt} shows a distribution of count rate of HXI2 top-layer DSSD with $PI>800$ corresponding to $E>80.0$~keV, which should have only background events. Clearly, the background level in top-layer DSSD is lower at $-50^\circ\lesssim lat.\lesssim100^\circ$ and $lon.\gtrsim 0^\circ$, while it is higher at $-150^\circ\lesssim lat.\lesssim-100^\circ$. In order to investigate the position dependence of the top-layer DSSD background, spectra are extracted from 6 regions, which are divided at $lon.=-60^\circ, 60^\circ$ and $lat.=0^\circ$. In Fig.~\ref{fig:highlowE}, spectra from regions with highest and lowest background levels are plotted. Although there is a large difference at higher energies, it does not affect the scientific analysis because signals above 30~keV in top-layer DSSDs are discarded in the energy-dependent layer selection. At energies below 30~keV, the background level can change by a factor of 3 at maximum depending on the orbital phase.

\subsection{Properties of CdTe-DSD Background}
NXB in the CdTe-DSD (Layer 4) of HXI is composed of many activation lines from radioactive isotopes induced by geomagnetically-trapped protons in the SAA as shown in the cyan histogram in Fig.~\ref{fig:screenedNXB}. Since orbit inclination angle of {\it Hitomi} is $31^\circ$, the HXI passes the SAA for 8--9 times a day. Low energy protons trapped in the SAA generate radioactive isotopes inside the detectors via interactions between protons and heavy atoms such as Cd, Te and Bi contained in the CdTe-DSDs and BGO shields. Gamma-ray photons and $\beta$ particles from these radioactive isotopes are the main cause of the CdTe-DSD background.

\begin{figure}[tbp]
\begin{center}
\includegraphics[width=0.8\hsize]{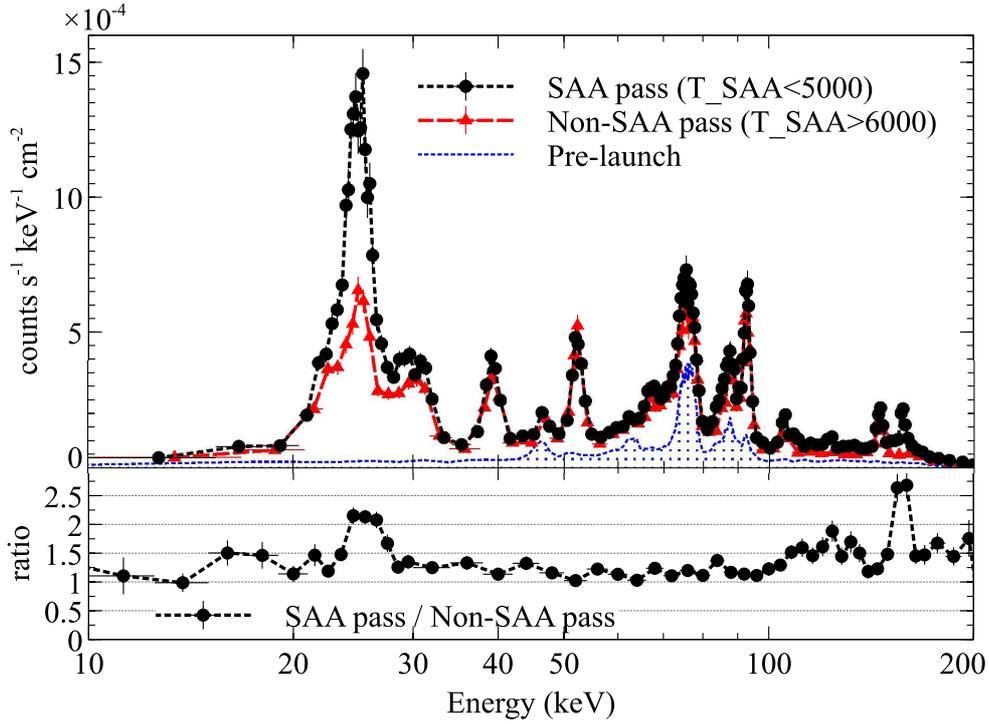}
\caption{HXI2 NXB spectra of CdTe-DSD sorted with T\_SAA. Lines at 46~keV and 63~keV in pre-launch data are thought to be intrinsic lines. The lines at $70\textrm{--}80$~keV are Bi-K lines from the BGO shield.}
\label{fig:CdTeSAA}
\end{center}
\end{figure}

Since the radioactive isotopes are generated in the SAA passage, the CdTe-DSD background depends on the time after the SAA passage, which is defined as {\tt T\_SAA}. Fig.~\ref{fig:CdTeSAA} shows spectra sorted by {\tt T\_SAA}. In this figure, the SAA pass is defined as {\tt T\_SAA<5000}, where $5000$~sec roughly corresponds to one orbital period of {\it Hitomi}, and non-SAA pass is defined as {\tt T\_SAA>6000}. It is clear that a few lines at $\simeq25$~keV and $\simeq160$~keV are rapidly decaying after the SAA passage, but spectra at 30--80~keV, which are used for scientific analysis, show less variability $\lesssim30$\%.

The properties of activation background in the CdTe-DSD observed by the HXI gives us essential information to understand the activation background, and significantly improved the accuracy of the simulations of the activation background. Details of the simulation studies are described in Odaka et al. 2017\cite{Odaka2017}. This result will be a great help for future hard X-ray missions such as {\it FORCE}\cite{Mori2016}.

\subsection{Final background spectra}
\begin{figure}[tbp]
\begin{center}
\includegraphics[width=0.8\hsize]{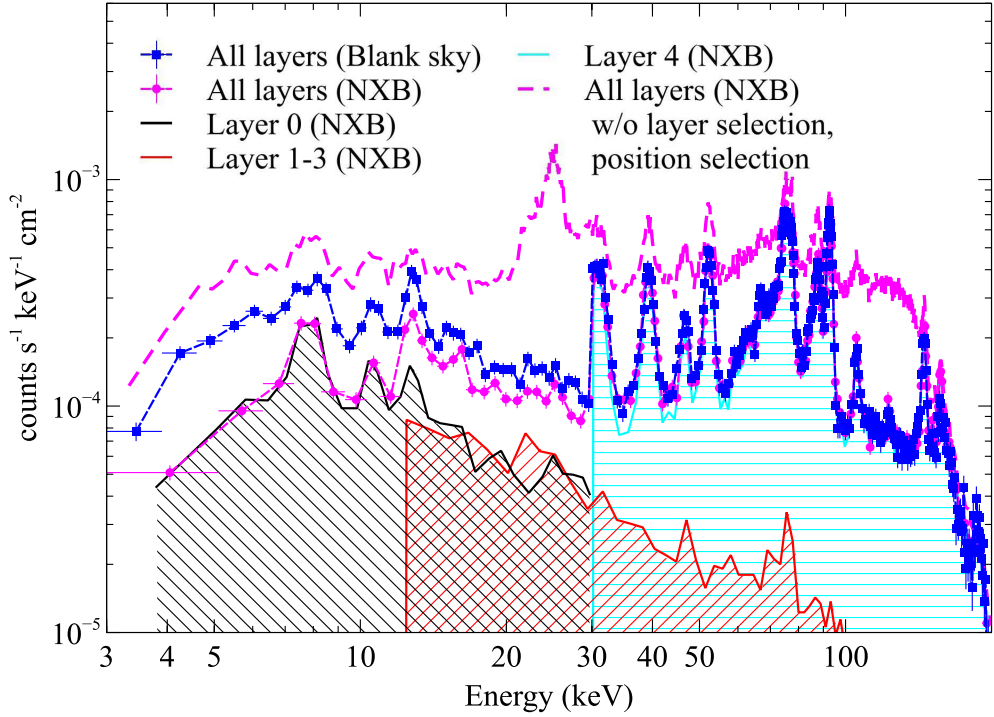}
\caption{The final spectrum of HXI2 NXB and blank sky. The spectra in Layer 0, Layer 1--3 and Layer 4 of HXI2 NXB after the energy-dependent layer selection are shown as shaded regions.}
\label{fig:finalnxb}
\end{center}
\end{figure}

The final spectrum of the NXB and blank sky after all the processing and screening is shown Fig.~\ref{fig:finalnxb}. Compared with the NXB spectra before the energy-dependent layer selection (Fig.~\ref{fig:screenedNXB}), the total background level is clearly reduced by ignoring the strong line at 20--30~keV in CdTe-DSD and the albedo electron component dominating the higher energy region in top-layer DSSD. The NXB level is as low as the pre-flight requirement of $1\textrm{--}3\times10^{-4}$~counts~s$^{-1}$~cm$^{-2}$~keV$^{-1}$. The photon detection efficiency of the top-layer DSSD above 30~keV is ignorable, while that of the 4 layers of DSSDs (in total 2-mm thick) below 30~keV is $\gtrsim50$\%. Therefore, the energy-dependent layer selection efficiently reduce the background with a small loss of detection efficiency. It demonstrates the effectiveness of the design of the stacked semiconductor detector for achieving better sensitivity.

\begin{figure}[tbp]
\begin{center}
\includegraphics[width=0.8\hsize]{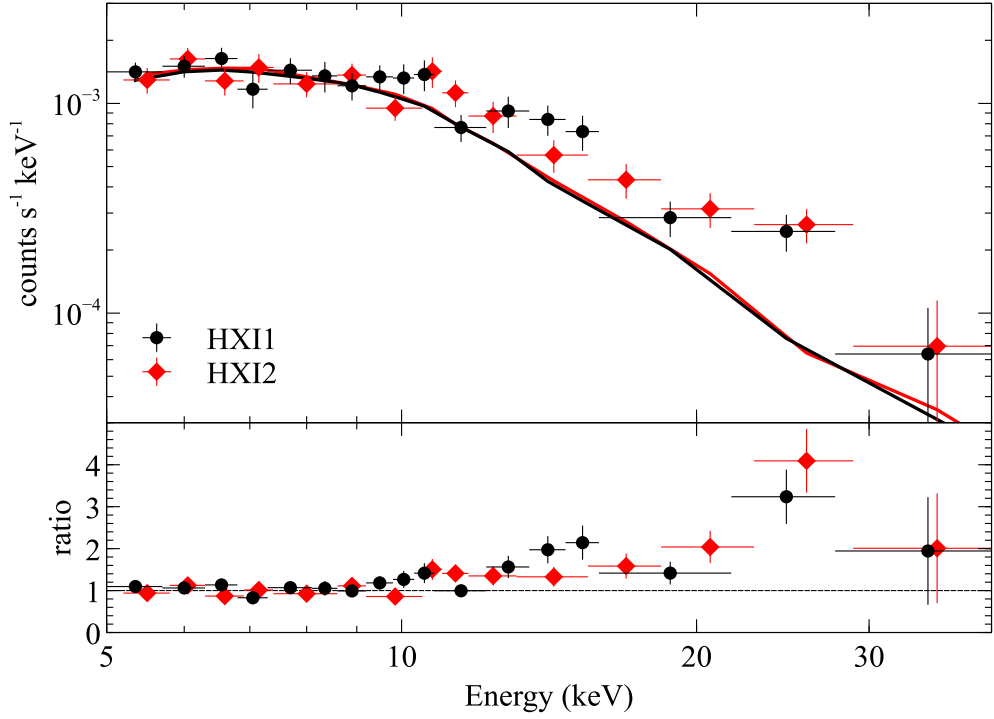}
\caption{The cosmic X-ray background (CXB) spectra observed by the HXI (black/red points) compared with the historically measured CXB spectral model\cite{DeLuca2004} (black/red slid curves).}
\label{fig:cxb}
\end{center}
\end{figure}

Fig.~\ref{fig:finalnxb} shows one additional demonstration of the HXI performance. A clear spectral difference between the blank sky and NXB spectra can be seen below $\simeq30$~keV. It indicates that the HXI is able to detect the cosmic X-ray background (CXB) below $\simeq30$~keV. Indeed, as shown in Fig.~\ref{fig:cxb}, the CXB spectra are significantly detected by the HXI. In this figure, a powerlaw model with a photon index of $\Gamma=1.41$ is overplotted as a historically measured spectral model of the CXB\cite{DeLuca2004}. Spectral fitting by this model with a fixed photon index provides the best-fit powerlaw normalization at 1~keV of $(9.0\pm0.5)\times10^{-5}$~cm$^{-2}$~photon~s$^{-1}$~keV$^{-1}$. Here, the spectral fitting is performed in 5--10~keV because the 15--30~keV spectra deviate from the powerlaw model due to the variability of the albedo electron background as previously shown in Fig.~\ref{fig:highlowE}. If the accuracy of background modeling is improved, it would be possible to detect the CXB spectra above 30~keV. A ratio of the best-fit normalization with the value reported by De Luca et al. (2004)\cite{DeLuca2004} is $1.13\pm0.06$, which is roughly consistent with the CXB fluctuation, $\sigma_{\rm CXB}/I_{\rm CXB}\simeq13$\% based on a relation of $\sigma_{\rm CXB}/I_{\rm CXB} \propto{\Omega_{\rm e}}^{-0.5}{S_{\rm C}}^{-0.25}$\cite{Nakazawa2009} with the HXI FoV $\Omega_{\rm e}=0.023$~deg$^{2}$ and an assumed upper cutoff flux $S_{\rm C}=8\times10^{-14}$~erg~s$^{-1}$~cm$^{-2}$. Thus, this result shows that the HXI has a good sensitivity for extended sources, which enables to detect the CXB below 30~keV.

\section{Conclusions} \label{sect:conc}
The HXI showed good performances and provided us important insights on the NXB in the hard X-ray band although it was lost after only two weeks of observations. The Crab spectra are well reproduced by the detector response constructed on the ground calibration data. The residual between the Crab spectra and the best-fit absorbed powerlaw model is less than $\lesssim5$\% at energies below 50~keV. The best-fit spectral parameters of the Crab are consistent with the historically reported values. This result indicates the correctness of the telescope effective area of the HXT and the detector response of the HXI. The NXB in top-layer DSSD is found to be dominated by the background due to low energy albedo electrons. Utilizing its strong dependence on the latitude and longitude, it can be reduced to 10--20\%. Even after this selection, the electron background in 20--30 keV varies by a factor of 3, depending on the orbital phase. The activation background in the CdTe-DSDs above 30~keV is more stable within $\lesssim20$\%. The final spectrum of the NXB after all the processing and screening satisfies the pre-flight requirement level of $1\textrm{--}3\times10^{-4}$~counts~s$^{-1}$~cm$^{-2}$~keV$^{-1}$, and it enables to detect the cosmic X-ray background. The properties of the in-orbit background of the HXI would be useful for the future hard X-ray missions.

\appendix    

\acknowledgments 
We acknowledge all the {\it Hitomi} team members, including many graduate students, for their great contributions to the HXI and the {\it Hitomi} project. We acknowledge support from JSPS/MEXT KAKENHI grant numbers 24105007, 15H03639, 25287059, 24244014, and the JSPS Core-to-Core Program. All U.S. members acknowledge support through the NASA Science Mission Directorate. Stanford and SLAC members acknowledge support via DoE contract to SLAC National Accelerator Laboratory DE-AC3-76SF00515 and NASA grant NNX15AM19G. French members acknowledge support from CNES, the Centre National d'Etudes Spatiales.


\bibliography{report}   
\bibliographystyle{spiejour}   


\vspace{2ex}\noindent\textbf{Kouichi Hagino} is an assistant professor at Tokyo University of Science. He received his BS and MS degrees in physics from the University of Tokyo in 2010 and 2012, respectively, and his PhD degree in physics from the University of Tokyo in 2015. He has been working on development of semiconductor detectors for high energy astrophysics.

\vspace{1ex}

\listoffigures
\listoftables

\end{spacing}
\end{document}